\documentclass{aastex}
\usepackage{emulateapj5}
\usepackage{onecolfloat5}
\let\gsim=\gtrsim

\makeatletter

\newenvironment{figurehere}{\begin{figure}[tb]\epsscale{1}}{\end{figure}}
\makeatother
\newcommand{\rad}{r}    

\long\def\comment#1{}

\def\W2{{\cal W}}

\newcommand{\deld}{\delta^{\rm D}}
\newcommand{\bn}{\hat{\bf n}}
\newcommand{\bm}{\hat{\bf m}}

\newcommand{\tableskip}{\\[-6pt]}

\newcommand{\Ylmn}{Y_{l}^{m}}
\newcommand{\alm}[1]{a_{l_#1 m_#1}}

\def\be{\begin{equation}}
\def\ee{\end{equation}}
\def\bea{\begin{eqnarray}}
\def\eea{\end{eqnarray}}

\def\Mpc{\,{\rm Mpc}}

\def\cmm2{{\,\rm cm^{-2}}}
\def\cm2{{\,{\rm cm}^2}}
\def\cmm3{{\,{\rm cm}^{-3}}}
\def\gcmm3{{\,{\rm g\,cm^{-3}}}}

\def\fun#1#2{\lower3.6pt\vbox{\baselineskip0pt\lineskip.9pt
  \ialign{$\mathsurround=0pt#1\hfil##\hfil$\crcr#2\crcr\sim\crcr}}}

\def\edge{{\it EDGE}}
\def\planck{{\it Planck}}
\def\GHz{\,{\rm GHz}}
\lefthead{KNOX ET AL.}
\righthead{}

\begin{document}
\twocolumn[
\submitted{To be submitted to ApJ}
\title{Probing Early Structure Formation with Far--Infrared
Background Correlations}
\author{Lloyd Knox, Asantha Cooray, Daniel Eisenstein\altaffilmark{1}}
\affil{Department of Astronomy and Astrophysics,
University of Chicago, Chicago, IL 60637, USA,
email: knox@flight.uchicago.edu, asante@oddjob.uchicago.edu,
eisenste@oddjob.uchicago.edu}
\and
\author{Zoltan Haiman\altaffilmark{1}}
\affil{Department of Astronomy, Princeton University, email:
zoltan@astro.princeton.edu}

\begin{abstract}  
The large--scale structure of high--redshift galaxies produces correlated
anisotropy in the far--infrared background (FIRB).  In regions of the sky where
the thermal emission from Galactic dust is well below average, these
high--redshift correlations may be
the most significant source of angular fluctuation power over a wide range of
angular scales, from $\sim\!7'$ to $\sim\!3^\circ$, and frequencies, from 
$\sim\!400$ to $\sim\!1000$~GHz.
The strength of this signal should allow detailed studies of the
statistics of the FIRB fluctuations, including the shape of the angular power
spectrum at a given frequency and the degree of coherence between FIRB maps at
different frequencies.  The FIRB correlations depend upon and hence constrain
the redshift--dependent spectral energy distributions, number counts, and
clustering bias of the galaxies and active nuclei that contribute to the
background.  We quantify the accuracy to which \planck\ and a newly proposed
balloon--borne mission \edge\ could constrain models of the high--redshift
universe through the measurement of FIRB fluctuations.  We conclude that the
average bias of high--redshift galaxies could be measured to an accuracy of
$\lesssim\!1\%$ or, for example, separated into 4 redshift bins 
with $\sim\!10\%$ accuracy.
\end{abstract}

 ]

\altaffiltext{1}{Hubble Fellow}

\section{Introduction}

The discovery of the cosmic far--infrared background
\citep[FIRB;][]{puget96,fixsen98,dwek98,schlegel98,lagache99} and the
determination of its spectrum have resulted in a new probe of structure
formation in the high--redshift universe
\citep[e.g.][]{guiderdoni98,blain99a,haiman00}.  A compelling explanation for
the background is that it results from the thermal emission of interstellar
dust associated with high--redshift galaxies and heated by the internal optical
and ultraviolet (UV) radiation from stars and to a lesser extent active
galactic nuclei (AGN) \citep{stecker77,bond86}.  Thus, the basic properties of the FIRB are sensitive to
the role of galaxy formation and subsequent evolution.

Observing the FIRB at higher angular resolution and higher sensitivity than was
done by the Far Infrared Absolute Spectrometer (FIRAS, \citet{mather99}) 
instrument on the Cosmic Background Explorer ({\it COBE}) 
satellite should reveal the presence
of correlated fluctuations resulting from the correlations in the galaxies
contributing to the background.  Haiman and Knox (2000, hereafter HK00) used
simplified semi--analytic models for the origin of the background flux to show
that these correlated anisotropies have an amplitude that is roughly 10\% of
the mean, a level detectable with current technologies.  The purpose of the
present paper is to (1) spell out and quantify what one could learn from
detailed observations of the FIRB correlations, (2) show that such detailed
studies are possible due to the absence of other strong sources of fluctuation
power in the relevant range of frequencies and angular scales, and (3) show
that FIRB anisotropy observations provide a powerful complement to direct
high--angular resolution observations of the individual sources.

As we discuss in detail later, the background at different frequencies is
composed of sources from differing ranges of redshifts.  FIRB maps at multiple
frequencies are therefore not expected to correlate perfectly with each other.
We show how the shape of the FIRB angular power spectrum at different
frequencies and the correlations between the maps can be used to determine
physical properties of contributing high--redshift sources, in particular the
product of the infrared emissivity of the sources and their bias relative to
the dark matter density field.  With assumptions about the source biasing and
spectra, the infrared emissivity as a function of redshift can be converted to
a measurement of the energy--production history of the universe, complementing
current approaches involving optical and UV observations
\citep[e.g.][]{madau97,madau98}.  Conversely, one could combine FIRB anisotropy
measurements with other measures of the emissivity density---e.g, from deep
sub-millimeter surveys followed up by redshift determinations---to determine the
bias of the sources.

Other important sources of fluctuation power over the relevant range of
frequencies and angular scales are the cosmic microwave background (CMB), the
shot noise due to the discrete nature of the FIRB sources, and thermal emission
from dust in our own Galaxy.  The CMB prevents measurements of the FIRB at
frequencies less than about 200~GHz.  Even at 200~GHz the angular power
spectrum of the CMB is over an order of magnitude larger than the 
FIRB's, but with a
sufficiently sensitive lower frequency CMB map one could use the well--known
spectral dependence of the CMB to subtract it with high precision.  We expect
the correlated FIRB fluctuations to dominate the shot--noise contribution at
multipole moments $\ell \la 1500$, or angular scales $\ga 7'$.  Thermal
emission from dust in our own galaxy could easily be confused with
FIRB anisotropy, but, as we show below, the dust fluctuation power has a
strong spatial dependence.  There are regions of sufficiently low
dust fluctuation power that the FIRB fluctuations are dominant
at $\ell \ga 60$ (angular scales $\la 3^\circ$) at low
frequencies.  This critical multipole moment slowly increases as the frequency
increases towards $\nu \sim 1000$~GHz and then rapidly increases beyond there,
as the Milky Way dust appears hotter than the high--$z$ dust contributing to the
FIRB.  Between these frequency and multipole moment limits, the FIRB
correlations can be studied without the need for aggressive foreground
subtraction.

Our work has been largely motivated by the proposed balloon--borne Explorer of
Diffuse Galactic Emissions\footnote{http://topweb.gsfc.nasa.gov}  
(\edge), which will survey the sky in 10
frequency bands from 150~GHz to 1290~GHz, with angular resolution ranging from
14$'$ to 6$'$ full--width half--maximum (FWHM).  Later in the decade, the 
{\it Planck}
surveyor\footnote{http://astro.estec.esa.nl/Planck/; also, ESA D/SCI(6)3.}
should provide high--quality maps of correlated FIRB anisotropy at 217, 320, 545
and 850~GHz.  We forecast the errors on the scale-- and redshift--dependent
product of bias and emissivity that could be reconstructed from \edge\ and
\planck\ observations.  We also discuss how these quantities could be
disentangled from uncertainties in the cosmological model and the emission
spectrum of the sources.

Both sets of observations can reconstruct the product of emissivity and bias to
$\sim\!10$\% in 4 redshift bins with the highest one extending from $z =2$ to
$z=4$.  Indeed, the constraints are $\sim\!1\%$ if one considers smoother
models for the bias--weighted emissivity.  This level of precision 
is hard to duplicate by directly
observing the clustering properties of the sources.  It may be possible to
resolve the background as a function of redshift in several frequency bands in
the next decade using, e.g., BOLOCAM \citep{glenn98} on the Large
Millimeter Telescope (LMT) or the 
Atacama Large Millimeter Array\footnote{http://www.alma.nrao.edu} (ALMA) 
and doing
follow--up optical redshift determinations.  However, we show that determining
the clustering of the sources to $\lesssim10\%$ accuracy at large enough scales
to be in the linear regime requires many thousands of sources with follow--up
redshifts of very faint galaxies.

Other missions may also detect the correlated FIRB
anisotropy.  The TopHat experiment\footnote{http://topweb.gsfc.nasa.gov/}, with
its 660~GHz channel, may provide an early opportunity for studying the FIRB
anisotropy.  The balloon--borne Bolometric Large Aperture
Sub-millimeter Telescope\footnote{http://www.hep.upenn.edu/blast/} 
({\it BLAST}), designed to detect sources at the bright and rare end
of the source count distribution, may also be able to measure 
fluctuations in the diffuse background.

Indeed, detectable FIRB correlations may even be lurking in existing
data sets.  Using the ISOPHOT instrument on ISO\footnote{see
http://isowww.estec.esa.nl/ for details}, the shot noise of the FIRB
has been seen at 170$\mu$m (1760~GHz) but the galactic dust emission,
at least in the field observed, has obscured the FIRB correlations
\citep{lagache00}.  The correlations may be detected upon further
analysis of the ISO data (after some dust cleaning), or in the {\it
BOOMERanG} 420 GHz data \citep{deb99}, or even possibly in the {\it
COBE}/FIRAS data.  \citet{kashlinsky00} claim a tentative detection of
correlated fluctuations in the {\it near}--infrared background 
using {\it COBE}/DIRBE data shortward of 10~$\mu$m, with somewhat
greater amplitude than that predicted by \citet{JimKas97}.

The remainder of our paper is organized as follows.  In \S~2, we
review what is known about the sources of the FIRB and about the mean
spectrum.  In \S~3, we describe both the correlated and shot--noise
contributions to the anisotropy.  For the correlated component we
demonstrate the dependence of the angular power spectrum on the
fluctuation power in the FIR emission at different redshifts.  In
\S~4, we describe the fluctuation power in the sky at the frequencies
of interest from the other relevant components: thermal emission from
dust in our own Galaxy, the CMB, and the FIRB.  In particular, we
emphasize the wide distribution in amplitudes of dust power spectrum
across the sky.  In \S~5, we discuss qualitatively the reconstruction
of various interesting physical quantities from FIRB observations, and
then forecast \edge\ and \planck\ errors on a reconstruction of a
scale-- and redshift--dependent product of bias times emissivity.  We
discuss our results in \S~6, with an emphasis on the complementary
nature of point source observations, and summarize our conclusions in
\S~7.

Although we maintain generality in all derivations, we illustrate our
results with the currently favored $\Lambda$CDM cosmological
model. Our choices for cosmological parameters are $\Omega_c=0.30$,
$\Omega_b=0.05$, $\Omega_\Lambda=0.65$, $h=0.65$, $n=1$, and {\it
COBE} normalization of $\delta_H=4.2 \times 10^{-5}$
\citep{bunnwhite97}.  We use the fitting formula for the transfer
function given by \citet{eishu99}.  This model has mass fluctuations
on the $8 h^{-1}$ Mpc scale in accord with the abundance of galaxy
clusters $\sigma_8=0.86$ \citep{vianna99}.

\section{FIRB Mean}

In this section, we review what is known about the FIRB mean, both from the
FIRAS determination of its spectrum and from observations of sources---most
notably at 850$\mu$m by the Sub--millimetre Common User Bolometer Array (SCUBA) camera
\citep{holland98} on the James Clerk Maxwell Telescope (JCMT). 
We emphasize the connection, created by the
redshifting of the peak of FIR emission, between studying the background
spectrum at low frequencies and determining the FIR emissivity at high
redshifts.

\subsection{The Spectrum}

A grey body with power--law emissivity has been fit to the spectrum
of the mean FIRB as determined with the {\it COBE}/FIRAS data
\cite[hereafter F98]{fixsen98}:
\be
 I_\nu = \tau_0 (\nu/\nu_0)^\alpha B_{\nu}(T)
\ee
where $B_{\nu}(T)$ is the Planck function and $\tau_0$ is the optical depth at
$\nu_0 = 3$~THz.  F98 find $\tau_0 = (1.3 \pm 0.4)\times 10^{-5}$, $T=18.5 \pm
1.2$K and emissivity power--law index $\alpha = 0.64\pm 0.12$; this spectrum is
plotted in Figure~\ref{fig:tfirb}.

\begin{figurehere}
\plotone{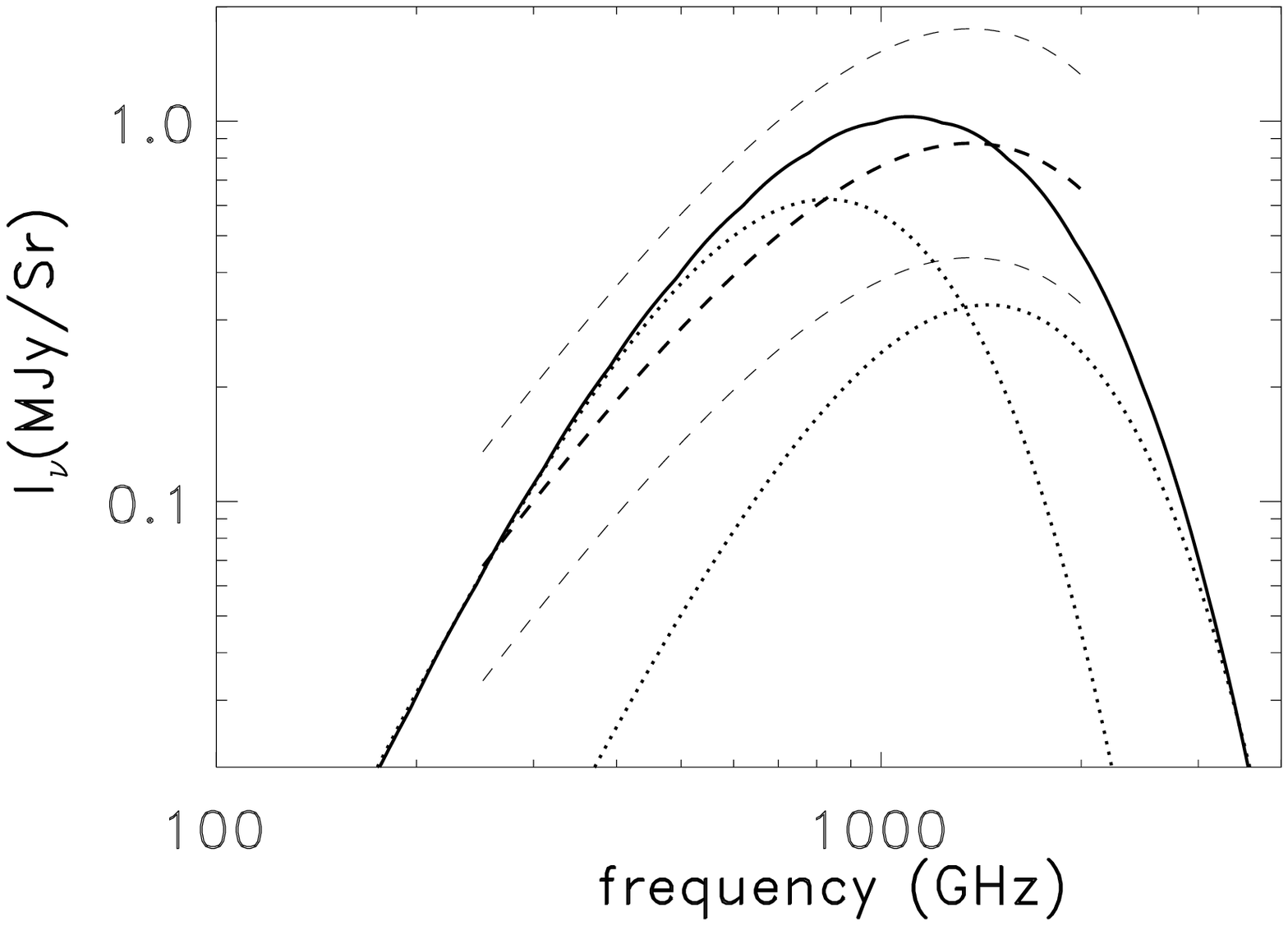}
\caption[]{\label{fig:tfirb} The spectrum of the FIRB predicted in the HK00
standard model (solid curve).  The dashed curves show the measurements with
$\pm 1\sigma$ uncertainties from F98.  The light dotted curves are grey--body
spectra with $\alpha=2$ and $T=8$~K (left curve) and $T=14$~K (right curve).
Their optical depths and temperatures are chosen to match the low frequency and
high frequency sides of the HK00 spectrum.  The HK00 spectrum is a sum of many
such grey--body spectra.  }
\end{figurehere}
%

HK00 considered a variety of models to describe the FIRB spectrum.  For the
present calculations, we use only the ``standard model'' from HK00, which
we refer
to simply as the HK00 model.  Its spectrum is shown in
Fig.~\ref{fig:tfirb}, with the amplitude reduced by 40\% relative to HK00 to be
in better agreement with the FIRAS data.  We apply the reduction throughout
this paper, thereby reducing the angular power spectra by a factor of 2.  The
HK00 model for the FIRB mean assumes that the UV emissivity at each redshift is
proportional to the star--formation rate (SFR) as estimated by \citet{madau99},
that dust production is proportional to the SFR, and that the dust has the
optical properties of the Draine and Lee (1984) model.  With the optical
properties and radiation backgrounds specified, the dust temperature follows.
The two proportionality constants for the UV emissivity and dust production are
set to produce a spectrum that is consistent with the F98 determination.

On the Rayleigh--Jeans side of the spectrum, Draine and Lee (1984) dust is a
grey body with emissivity index $\alpha=2$.  Thus the HK00 spectrum is a sum of
such grey bodies with varying apparent temperatures and optical depths, where
apparent temperature is defined as $T_{\rm app} = T/(1+z)$ where $T$ is the
physical temperature.  The HK00 spectrum cannot be fit by a single grey body
with $\alpha=2$.  We demonstrate this in Fig.~\ref{fig:tfirb} with the
$\alpha=2$ grey bodies plotted as dotted curves.  For the right--most dotted
curve, we chose the temperature and optical depth to fit the curve at high
frequencies.  Moving towards lower frequencies, this $T_{\rm app} = 14$~K grey
body quickly falls below the HK00 spectrum.  To fit the spectrum, one must add
in colder components, such as the $T_{\rm app}=8$~K one shown.  Clearly, in
this particular model, there is not much need for even colder components.
Since the HK00 model has dust with $T \approx 24$~K in the redshift range 1 to
4, this indicates that most of the FIRB in this model comes from $z < 2$.  Note
that the F98 fit to the FIRAS data is even softer than the HK00 model, possibly
due to the presence of even colder (higher redshift) components, although
it should be kept in mind that the uncertainties in this slope at low 
frequencies are quite large.

In principle, one can take the spectrum of the background and, assuming the
emissivity index, reconstruct the optical depth of each temperature component.
This is the essence of the idea recently discussed and implemented by
\citet{gis00}.  By assuming the shape of the average SED, one can recover the
FIR luminosity density as a function of redshift, which can be related
to the SFR using certain assumptions.

Due to the large size of the FIRAS uncertainties, especially 
at frequencies less than
$\sim\!300$~GHz, this reconstruction can not be done accurately.
Furthermore, there are no proposed missions for improving upon FIRAS's
determination of the FIRB spectrum, which we understand to be a very difficult
task.  Fortunately, one can make quite similar use of a measurement of the
spectrum of the FIRB {\it anisotropy}---although with the added complication
that one is not reconstructing the luminosity density, but the luminosity
density weighted by the clustering bias of the sources.  A differential
measurement of the spectrum should be able to measure to lower frequencies due
to the fact that the amplitude of the CMB anisotropy is about $10^{-5}$ of the
CMB monopole.  This is especially interesting since the limiting redshift
(coldest component) one can reach is determined by the lower frequency limit of
the spectrum determination.  We turn to a description of the FIRB anisotropy in
\S~3.

\subsection{Resolving the Background}

The mean brightness of a background due to point sources depends on 
the number density of sources as a function of limiting flux, $N(>\!S)$:
\be\label{eq:sourcesmean}
I = \int (S^2 dN/dS) d\ln S.
\ee
Notice that the bulk of the background is contributed by sources near the
maximum of $S^2 dN/dS$.  We therefore show this function as predicted by the
semi--analytic modeling of Guiderdoni et al. (1998, hereafter G98) at several
frequencies in Fig.~\ref{fig:s2dnds}.  G98 calculated predictions for a range
of models; we use their model E predictions, which are the ones in best
agreement with the observations available at $850\mu$m.

\begin{figurehere}
\plotone{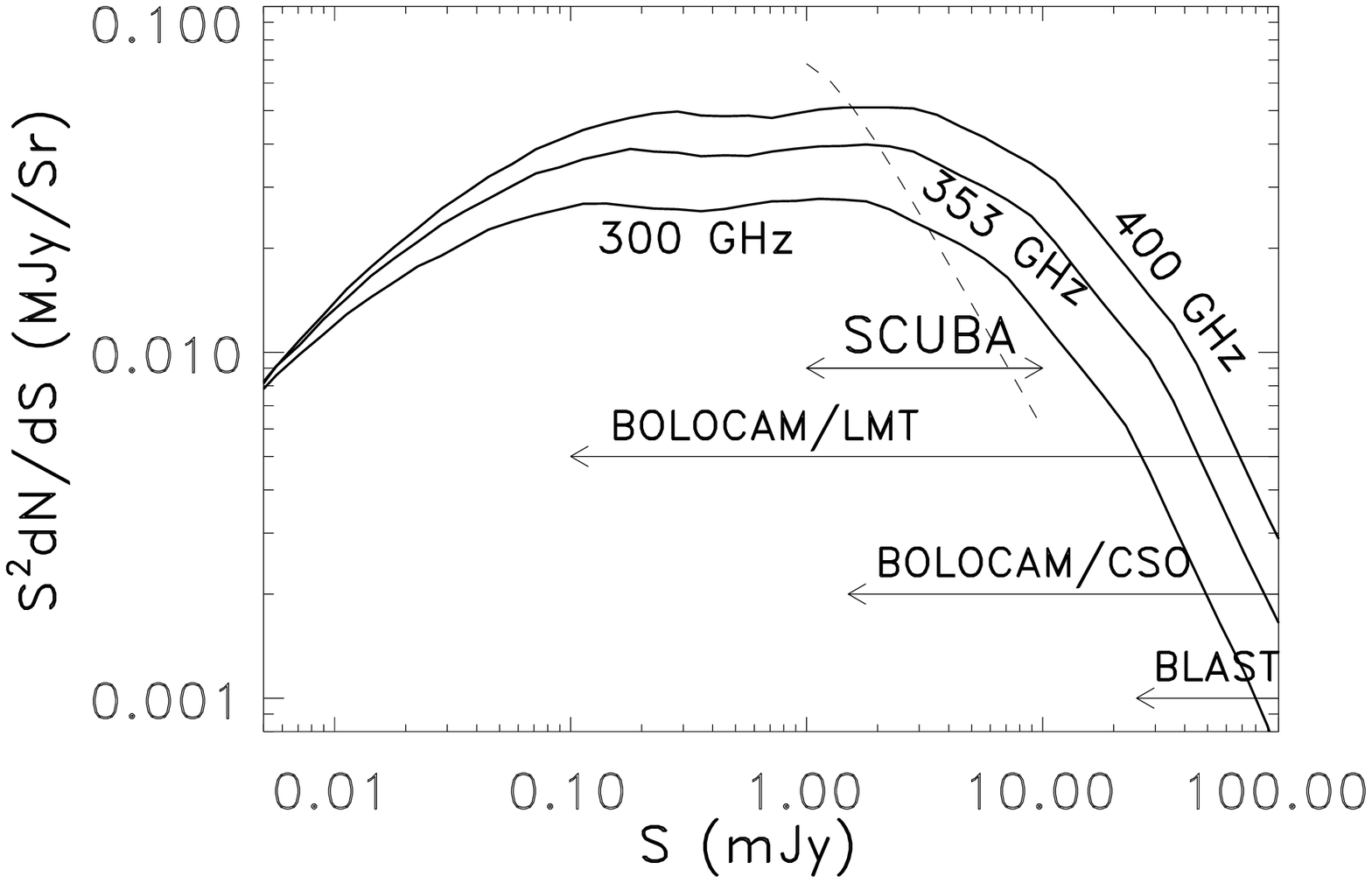}
\caption[]{\label{fig:s2dnds} The contribution to the FIRB mean from each
logarithmic interval in $S$, according to model E (G98) at 300~GHz,
353~GHz and 400~GHz.  The light dashed line is a fit to the SCUBA
data, which covers the $\sim\!1$~mJy to $\sim\!10$~mJy range
\protect\citep{barger99}.  Future observations near these frequency
ranges will go deeper and broader.  The left--hand limits for BOLOCAM,
observing at 270~GHz, are the 5$\sigma$ sensitivities after a 10--hour
pointing at a single field--of--view.  Both of these pointing should
result in the detection of about 60 sources according to model E
(G98).  The left--hand limit for BLAST is the $5\sigma$ limit for the
deepest of their planned surveys, which takes 2 days of a 10--day
long--duration balloon flight.  The vertical location of the indicated
survey ranges is arbitrary.  }
\end{figurehere}
%

Considerable progress has been made towards resolving the FIRB into
discrete point sources \citep{hughes98,barger98,eales99,smail97,
holland98,blain99b,barger99,puget99,barger99b,blain99c}.  In particular,
the SCUBA camera \citep{holland98} on the JCMT has been
used to identify point sources at $850\mu$m that account for a large
fraction, $\sim 50\%$, of the FIRB at these wavelengths \citep{barger99}.
As shown in Figure~\ref{fig:s2dnds}, the SCUBA detections are in the
range of $\sim\!1$~mJy to $\sim\!10$~mJy, presumably near the peak
of $S^2 dN/dS$.  The result of fitting a double power--law form for
$N(>\!S)$ to the SCUBA data, with the constraint that it reproduce the
FIRAS--determined mean \citep{barger99}, results in the light solid line
in Figure~\ref{fig:s2dnds}.

Followup observations of SCUBA locations at radio, optical and
infrared wavelengths have suggested possible counterparts, albeit with
a relatively low identification rate.  Typical sources are identified
as galaxies with moderate to massive star formation rates.  Optical
redshifts, the sub--millimeter to radio spectral index
\citep[e.g.][]{carilli99} and the detection of CO molecular lines
\citep{frayer99} suggest that most of these sources lie at redshifts
ranging from 1 to 4.  For a recent review of SCUBA results see
\citet{smail00}.  Continued followup observations of SCUBA surveys and
those that were carried out with ISO, such as the European Large Area
ISO Survey \citep{efstathiou00,serjeant00,oliver00}, will provide
important details on the physical properties of far--infrared (FIR)
emission and the redshifts of contributing sources.

Further progress towards resolving the background in wave bands near
850$\mu$m (353~GHz) will come soon from a bolometer array sensitive to
1100$\mu$m (273~GHz) called BOLOCAM, now at the Caltech Submm Observatory
(CSO).  When deployed on the LMT, it will
have sufficient resolution to resolve considerably more of the background.
{\it BLAST} will be able
to detect the brightest sources at $750\mu$m (400~GHz) and also has
detectors at $300$ and $200\mu$m, to permit the determination of color
redshifts.  The California Millimeter Array (CARMA), an interferometer
composed of existing BIMA and Owens Valley telescopes, is expected to
carry out observations at 235 and 345 GHz with $\sim$ arc second resolution that
will allow cross-identification at optical wavelengths, allowing
further followup observations \citep{carlstrom00}.  Finally, 
ALMA will have high sensitivity and very high angular
resolution.  In a single 10 hour pointing, it will achieve a 5$\sigma$
noise level of 0.03 mJy and still be far from the confusion limit.
However, many such pointings will be required to get large numbers of
sources, as the field--of--view is only 0.07 square arcminutes, over
which model E predicts about six sources brighter than 0.03~mJy.

At shorter far--infrared wavelengths, a smaller fraction of the background has
been resolved.  For example, the FIRBACK survey at 170 $\mu$m with ISOPHOT on
ISO \citep{puget99} has resolved about 10\% of the background \citep{puget99}.
Much progress will come in the near future from MIPS on {\it SIRTF} (160, 70
and 24$\mu$m), SPIRE and PACS on {\it FIRST} (90 to 500$\mu$m), and HAWC on
{\it SOFIA} (60, 110 and 200 $\mu$m).  Information on most of these
sub-millimeter and far--infrared instruments is summarized in \citet{blain99d}.

\section{FIRB Anisotropy}

The FIRB is anisotropic at small scales due to the discrete nature
of the sources, and at larger scales due to correlations between these
sources.  We first discuss the correlated anisotropy.

\subsection{FIRB correlations}

The antenna temperature of the FIRB at a given frequency $\nu$ and
in a given direction $\hat {\bf n}$ can be written as a line of sight integral
of the product between the mean FIR emissivity and its fluctuation:
\bea
T(\bn,\nu) = \int dz \frac{d\rad}{dz} a(z) \bar j(\nu,z)
        \left[1 + \frac{\delta
j(r(z)\bn,\nu,z)}{\bar j(\nu,z)}\right] \, ,
\label{eqn:firbmap}
\eea
where $\rad$ is the coordinate distance (or conformal time) from our
location at the coordinate origin and $\bar
j(\nu,z)$ is the mean emissivity per comoving unit volume at frequency $\nu$ as
a function of redshift $z$.  In a flat universe, which we will assume, $\rad$
is simply the proper motion distance.  We will always use comoving wavenumbers and spatial scales.

We next wish to derive the angular power spectrum $C_\ell^{\nu \nu'}$ of the
FIRB, which is the Legendre transform of the two--point correlation function
\begin{eqnarray}
C(\bn,\bm) &\equiv& \langle T(\bn,\nu) T(\bm,\nu') \rangle  \nonumber\\
       &=& \sum_{\ell} {2l+1\over4 \pi} C_l^{\nu \nu'} P_l(\hat{n}\cdot\hat{m}).
\label{eqn:twopoint}
\eea
This is most easily done by first decomposing the temperature maps
into spherical harmonic multipole moments, 
\begin{equation}
a_{lm}(\nu) = \int d\bn T(\bn,\nu) \Ylmn {}^*(\bn)\,.
\label{eqn:decompose}
\end{equation}
With an isotropic random field, the second moments of these 
multipole coefficients are
\begin{eqnarray}
\langle \alm{1}^*(\nu) \alm{2}(\nu')\rangle = \deld_{l_1 l_2} \deld_{m_1 m_2}
        C_{l_1}^{\nu \nu'}.
\end{eqnarray}
Using equation (\ref{eqn:firbmap}), we can rewrite equation (\ref{eqn:decompose}
as 
\begin{eqnarray}
\label{eqn:alm}
a_{lm}^{\rm FIRB}(\nu) &=& \int d\bn \Ylmn{}^*(\bn) T(\bn,\nu)
\nonumber \\
&=& i^l \int \frac{d^3{\bf k}}{2\pi^2} \delta({\bf k}) I_l^{\rm
FIRB}(k,\nu) \Ylmn{}^*(\hat{\bf k}) \, ,
\end{eqnarray}
where
\begin{eqnarray}
I_l^{\rm FIRB}(k,\nu) &=&  \int dz \frac{d\rad}{dz} W^{\rm FIRB}(k,\nu,z)
j_l(k\rad) \nonumber \\
        &\approx& W^{\rm FIRB}(k,\nu,z) \int dz \frac{d\rad}{dz}
j_l(k\rad) \nonumber \\\
        & = & W^{\rm FIRB}(k,\nu,z) \frac{\sqrt{\pi}}{2k}
\frac{\Gamma[(l+1)/2]}{\Gamma[(l+2)/2]} \, ,
\end{eqnarray}
and
\begin{equation}
W^{\rm FIRB}(k,\nu,z) = a(z) G(z) \bar j(\nu,z)
b(k,\nu,z) \, .
\end{equation}
Here, $G(z)$ is the linear theory growth function \citep{peebles80},
$j_\ell$ is the usual Bessel function, and the approximation
used follows the well--known Limber form \citep{limber53}. The ratio
of Gamma functions goes to $\sqrt{2/l}$ for $\ell \gg 1$.

In equation (\ref{eqn:alm}), $\delta({\bf k})$ is the linear
theory dark matter density fluctuation today, $\delta \equiv \delta \rho /
\bar \rho$.  It enters because we have assumed, as in HK00, 
that the fluctuations in the emissivity, $\delta j /\bar j$, 
are a biased tracer of those in the mass.  
In general, this bias
can depend on scale, frequency, and redshift, so that in Fourier space we
write:
\begin{equation}
\frac{\delta j({\bf k},\nu,z)}{\bar j(\nu,z)} = b(k,\nu,z) \delta({\bf k},z).
\end{equation}
We ignore the possibility of stochastic bias \citep{dekel99} until
the discussion.

With these assumptions about the bias we can 
finally write the angular power spectrum of FIRB anisotropy using the
three--dimensional, linear--theory power spectrum 
of dark matter density fluctuations today, $P_M(k)$,
\begin{equation}
\langle \delta({\bf k},z) \delta^*({\bf
k'},z) \rangle = (2 \pi)^3 \delta^D({\bf k} - {\bf k'}) P_M(k)G^2(z)
\, ,
\end{equation}
as a line of sight projection
\bea
\label{eqn:Limber2}
C^{\nu \nu'}_l &=& \int \frac{dz}{r^2}\frac{d\rad}{dz} a^2(z)
j_b(k,\nu,z)
j_b(k,\nu',z) P_M(k)|_{k=\frac{l}{r}}G^2(z). 
\eea
We have used the Limber approximation which sets $k=l/r$.  
At $\ell=30$ the Limber approximation  is valid here 
to within 10\% of the exact (first order) calculation and 
rapidly converges to the exact value as $\ell$ increases.  
Since the bias factor enters into the expression for $C_l^{\nu \nu'}$
multiplied by the mean emissivity, we define the ``bias--weighted emissivity''
$j_b$ as
\be\label{eqn:jbdef}
j_b(k,\nu,z) \equiv b(k,\nu,z)\bar j(\nu,z) \, .
\ee

We do not use the above equations in their full generality.  We follow HK00 in
assuming that the bias is a constant, i.e., independent of frequency, redshift
and scale.  We take this bias to be 3, which will roughly match the
observations of the clustering of $z=3$ Lyman--break galaxies
\citep{steidel98,giav98,adel98}.  Bias is likely to increase with redshift
\citep{baugh99,blanton00}, and hence this approximation may down--weigh the
relative importance of fluctuation power at high redshift.  The bias may also
be frequency dependent, which could result from two different populations of
sources, each with different spectral energy distributions and clustering
properties.  We briefly discuss such a possibility below but neglect it for
now in order to illustrate the nature of the observables in the simplest
possible case.

We wish to understand how measurements of $C_l^{\nu \nu'}$ can be used to
determine the bias--weighted emissivity $j_b$ as a function of redshift.  To
that end we display contributions to the angular
power spectra from various redshift ranges in Figure~\ref{fig:dcldlnz}.  
As stated above, we have assumed
that the bias is redshift--independent and that the mean emissivity is that of
the HK00 standard model.  

Figure~\ref{fig:dcldlnz} shows that the relative contributions from the various
redshift ranges depends on the observed frequency and on the shape of the
matter power spectrum.  In the top panel, we consider a power--law power
spectrum, $P(k) \propto k^{-2}$, and show the shape of the redshift breakdown
$d C_l^{\nu \nu}/dz$ for three frequencies, $\nu = 210, 353$ and $1090$~GHz.
These are the integrands of equation (~\ref{eqn:Limber2}).
Note that lower frequencies probe to higher redshifts, as expected.  For any
power--law power spectrum, the shape of $d C_l^{\nu \nu}/dz$ as a function of
$z$ is the same for all values of $\ell$.

The non--power--law nature of physical power spectra lead to different shapes of
$d C_l^{\nu \nu}/dz$ for different values of $\ell$.  In the lower panel of
Figure~\ref{fig:dcldlnz}, we plot three different $\ell$ at fixed frequency for
the $\Lambda$CDM model.  Higher $\ell$ map to smaller scales, where the
spectral index of the power spectrum is more negative.  This means that higher
redshifts, where the spatial scale corresponding to a given $\ell$ is larger,
contribute more to the anisotropies.  Hence, higher $\ell$ probe higher
redshifts.


\begin{figurehere}
\plotone{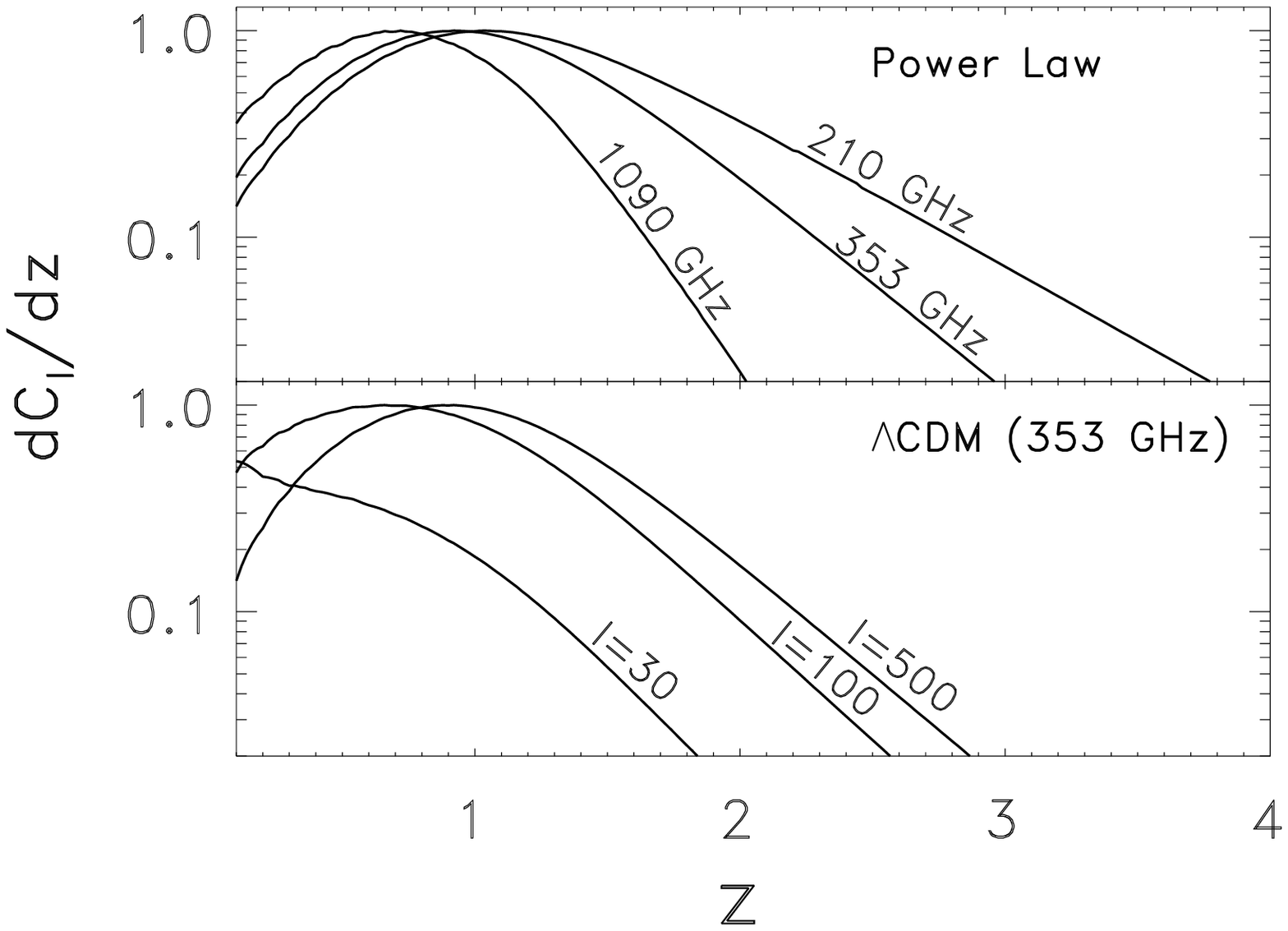}
\caption[]{\label{fig:dcldlnz} $dC^{\nu \nu}_l/dz$ for a power--law 3D power
spectrum (solid lines) at, from right to left, $\nu=$ 210, 353 and 1090~GHz,
all normalized to unity at their maximum. For any power--law 3D power spectrum,
these are independent of $\ell$.  The three curves in the bottom panel show the
same quantity, using a $\Lambda$CDM power spectrum at 353~GHz, for $\ell=30$,
100, and 500 respectively. }
\end{figurehere}
%

Figure~\ref{fig:dcldz} shows the contributions to $C_l^{\nu \nu}$ for different
redshift ranges and frequencies.  In each panel, one can see the shape
dependence of the contribution from each redshift interval, with lower
redshifts having more low--$\ell$ fluctuation power compared to the high--redshift
intervals.  This shape dependence aids in the reconstruction of the redshift
dependence of the fluctuation power.  Comparing the panels, one should note the
more rapid drop--off towards higher redshift at 850~GHz than at 210~GHz.  At
the higher frequency, there is very little contribution from $z\ga 1.5$.


\begin{figurehere}
\plotone{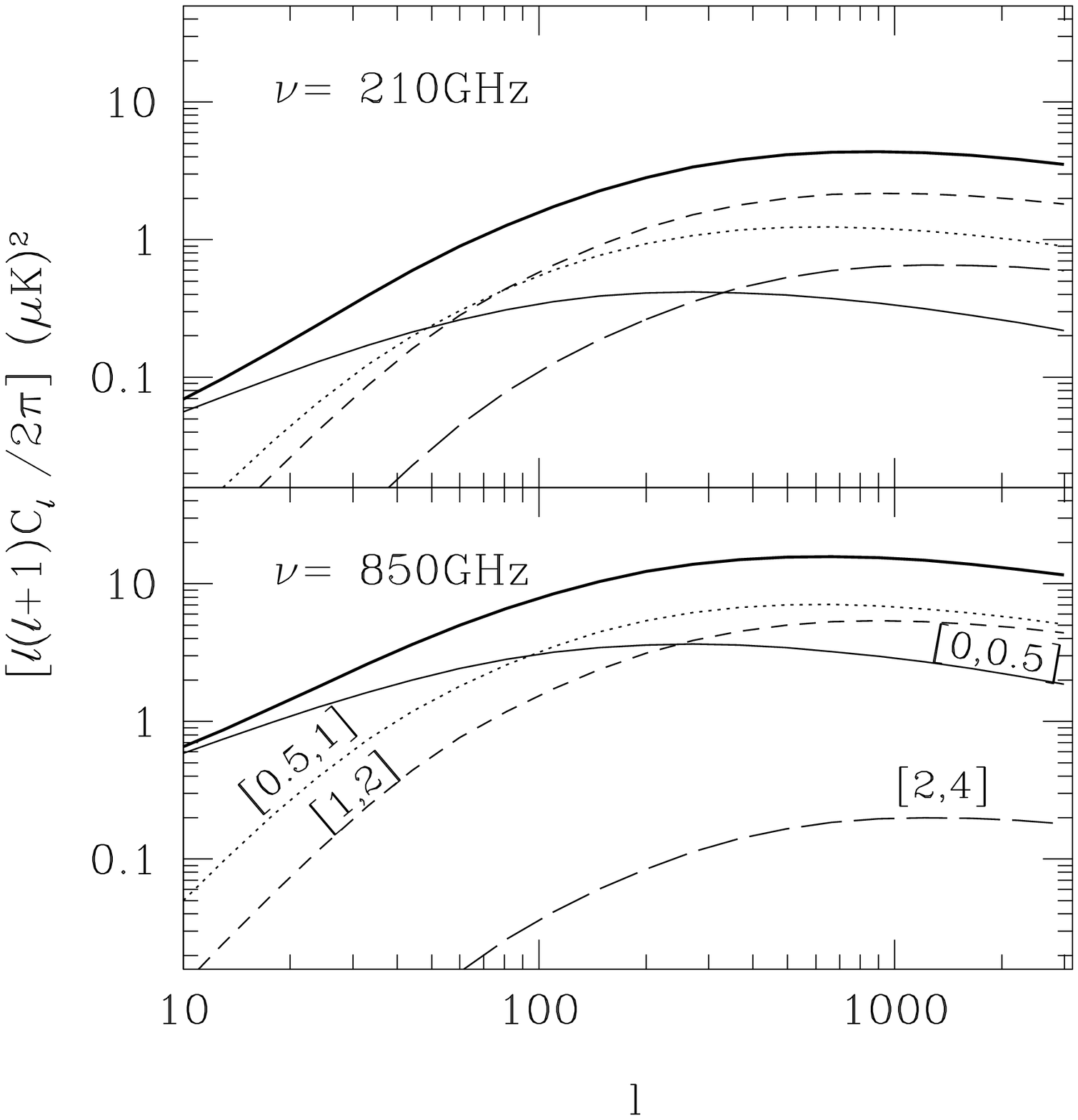}
\caption[]{\label{fig:dcldz}
Contributions to $C_l^{\nu \nu}$ from several redshift ranges (labeled
in lower panel) for $\nu = 210$~GHz (upper panel) and $\nu = 850$~GHz
(lower panel).  The heavy solid line is the total power spectrum
observed at $z=0$.}
\end{figurehere}
%

We now turn to the cross--frequency correlations.  In Figure~\ref{fig:cohere},
we show the expected cross--correlations between measurements at 390 GHz 
and those at four other frequencies (210, 852, 1060, and 1290~GHz). 
The lowest frequencies
are affected by all the redshifts that have any significant fluctuation power;
therefore, these frequencies weight all of the fluctuations in the same way and
are therefore highly correlated.  Higher frequencies are insensitive to
high--redshift fluctuations because they fall on the Wien side of the redshifted
spectrum.  With only a portion of the fluctuations included, the degree of
correlation degrades.  The break frequency, at which this correlation departs
from unity, is sensitive to the onset of significant fluctuation power in FIR
galaxies.  Because higher $\ell$'s are preferentially more sensitive to higher
$z$, the diminishing importance of high--$z$ at higher frequencies is more
pronounced at smaller angular scale.

It is important to keep in mind that 
the redshift--dependences of the power spectra, illustrated in
figures 3 and 4, and the cross--correlations shown in figure 5
are those of a particular model---the HK00 standard model.
For this model the emissivity tracks the star formation rate as determined by
\citet{madau98}, which is quite uncertain.  It is possible
that the SFR is much higher at high--redshift.  If one
assumes efficient conversion of gas into stars in all
collapsed halos down to some small mass (with virial temperatures
of abut $10^4$~K) then there is a peak in the SFR at $z \sim 7$
\citep[e.g.][]{haiman00b} which is seen in simulations
\citep{gnedin96}.
Increased high--$z$ SFR would lead to a greater fraction
of the fluctuation power coming from high redshifts, and greater decorrelation
between the high and low frequencies.

\begin{figurehere}
\plotone{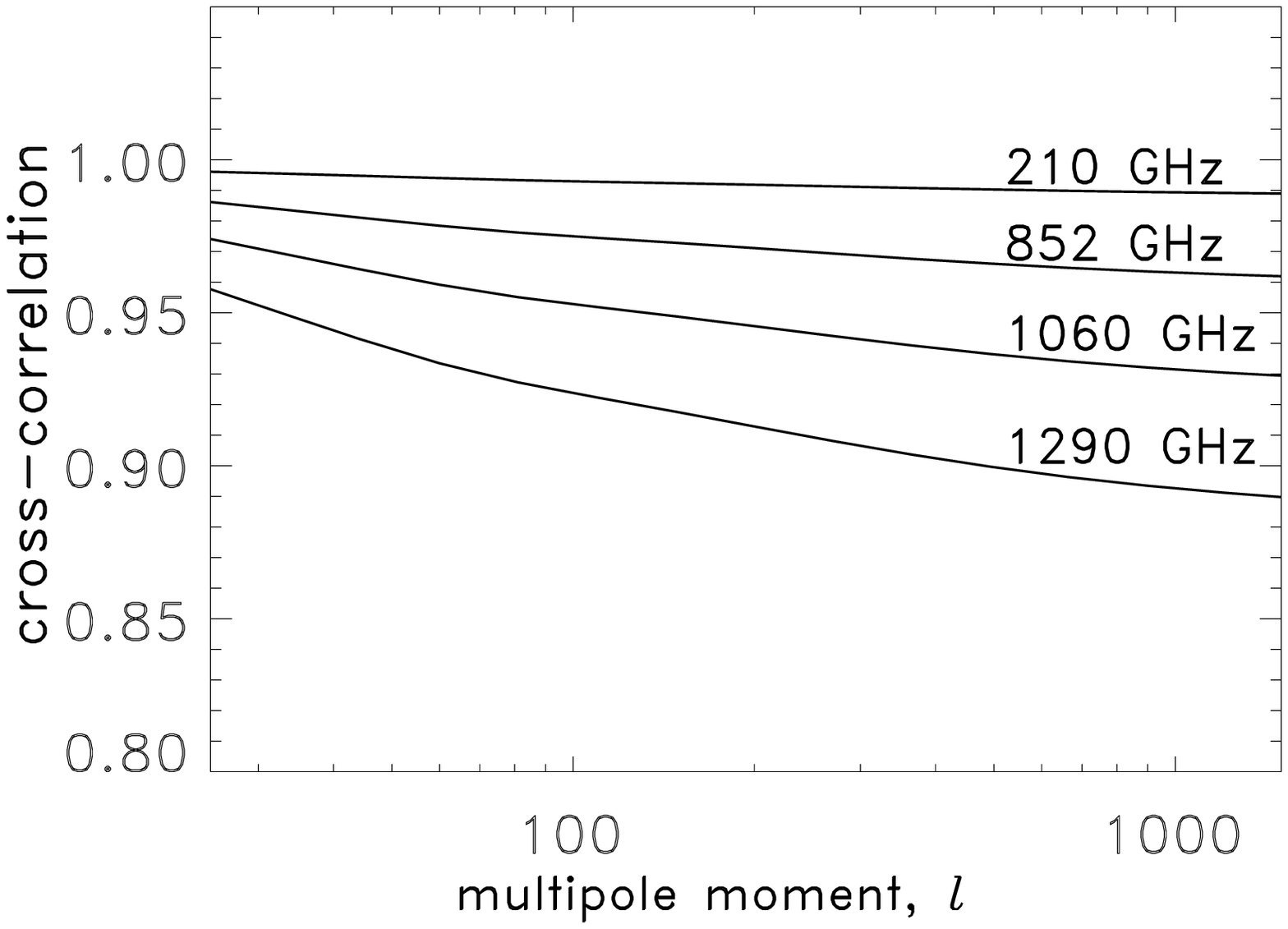}
\caption[]{\label{fig:cohere}
Cross--correlations $C_{\ell}^{\nu \nu'} / \sqrt{C_l^{\nu \nu}
C_l^{\nu' \nu'}}$ vs. $\ell$ between the $\nu = 390$~GHz channel and
four other frequencies $\nu' = $~210, 852, 1060, and 1290~GHz.  }
\end{figurehere}
%

We use the linear--theory evolution of the matter power spectrum in
our calculations.  At $z\approx1$, the non--linear corrections become
important at wavenumbers $k\gtrsim 0.4h\Mpc^{-1}$, which projects to
$\ell\gtrsim800$.  At lower redshifts, this non--linear scale moves to
smaller $\ell$, but the FIRB contributions drop as well (see the
higher $\ell$ curve in Fig.~\ref{fig:dcldlnz}).  At $\ell\approx
1000$, non--linear corrections produce a 20\% increase in the FIRB
angular power spectrum.  Because this is a small difference that can
be accurately included for any chosen cosmology, neglecting the
non--linearity won't affect our quantitative results.  However, we do
restrict our analysis to $\ell<1600$ both to avoid the deeply
non--linear regime and to eliminate the need for detailed shot noise
subtraction.

\subsection{Shot Noise}

At smaller angular scales, the fact that the FIRB comes from individual,
well--separated sources introduces anisotropy simply due to the Poisson 
sampling
of the density field.  This so--called shot noise can be calculated from the
source counts by
\be
C_l^{\rm shot} = \int (S^3 dN/dS) d\ln S.
\ee
Note that this is more heavily weighted towards the bright end of the
distribution than the mean brightness was (eq.~[\ref{eq:sourcesmean}]).

For a Euclidean distribution (appropriate for galaxies at low redshift), $dN/dS
\propto S^{-2.5}$ and so the integrand diverges at the bright end.  However, in
the case of the FIRB (at least as modeled by G98 model E), the enormous number
of faint, high--redshift sources cause the source counts to become steeper than
$S^{-2.5}$ at fainter flux levels.  Even if we remove sources only down to the
very conservative flux level $S_{\rm cut}$ such that $N(>\!S_{\rm cut}) = 1$
per $4\pi$ steradians, the FIRB shot noise is still dominated by the fainter,
high--redshift population.  Therefore, we take our shot noise estimates from the
G98 model E counts, assuming only a cut at one source per sky and avoid any
complications of source removal.  The number of sources detected at greater
than $5\sigma$ by \edge, according to model E, are given in
Table~\ref{table:channels}.  The removal of these sources from the map does not
significantly reduce the shot noise.

\section{Foregrounds and Backgrounds}

\begin{figure*}[bt]
\centerline{\epsfxsize=18cm\epsffile{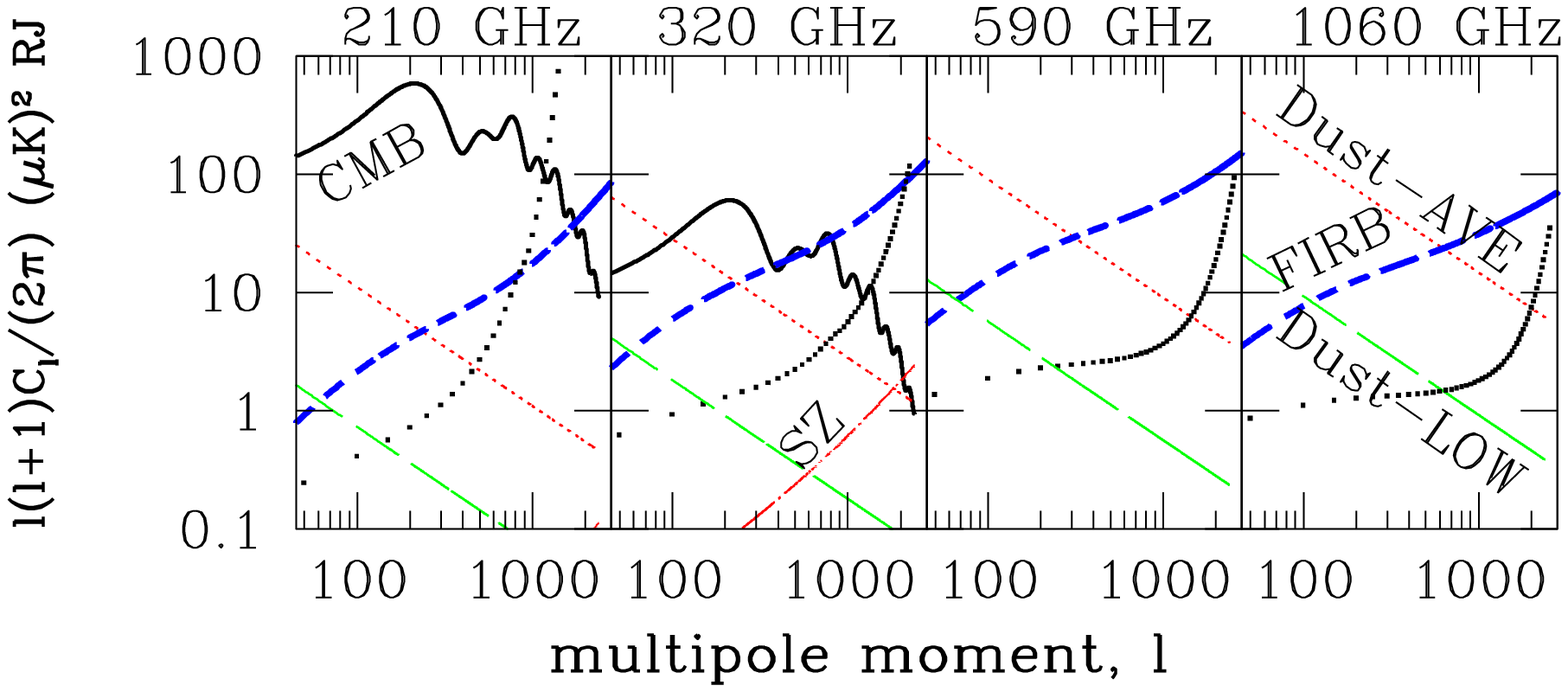}}
\vspace*{24pt}
\caption{\label{fig:fgnds} 
Power spectra of the dominant components of the far infrared sky
at four different frequencies measured by \edge.  All are labeled in at least
one of the panels.
The thick solid curves show the CMB in a flat $\Lambda$ CDM model; the thick
dashed curves show the FIRB predictions for the HK00 model including
contributions from nonlinear evolution and from shot noise as predicted
by model E (G98); the
thin dotted and long dashed curves show the dust spectrum representing an
average over the cleanest 50\% of the sky, as well as a spectrum of the
cleanest regions, respectively; and the thick (square) dotted curves with
spacings of $\delta l = 50$ are errors expected on the determination of the
FIRB power spectrum from a 10--day flight of \edge.}  
\end{figure*}

There are celestial sources of emission in the far infrared and
sub--millimeter other than the FIRB.  To detect
the correlations of the high--redshift galaxies that make up the FIRB, we need
to be able to separate this signal from all the other sources.  The two most
important contributions come from the CMB anisotropies and thermal emission
from dust in our own Galaxy.  Radio foregrounds such as synchrotron,
Bremsstrahlung, and rotational emission from dust do not appear to be
significant at $\nu\gsim200\GHz$.  Reviews of what is known about
galactic dust emission and other CMB foregrounds can be found in 
\citet{deoliveiracosta99}.

Figure~\ref{fig:fgnds} shows the predicted power spectra of the
components at 4 different \edge\ frequencies.  The FIRB angular power
spectrum is shown by the thick dashed curves for the standard,
constant bias model of HK00.  Unlike in all previous figures in this
paper, these curves include a shot--noise contribution, computed from
the G98 model E, as well as the effects of the non--linear evolution of
the matter power spectrum.  Other lines show the contributions of
the CMB and Galactic dust, which we will now discuss in detail.

\subsection{CMB}

The thick solid curves in Figure~\ref{fig:fgnds} show the CMB power spectrum in
our flat $\Lambda$CDM model, consistent with the {\it COBE}/DMR, {\it
Boomerang} and {\it Maxima} measurements \citep{debernardis00,hanany00,jaffe00}.  As one can see from this figure, the
CMB anisotropies are the dominant source of fluctuation power on the sky over a
substantial range of $\ell$ at low frequencies.  The FIRB anisotropies would be
swamped by such a signal.  Fortunately, the spectrum of the CMB anisotropies is
well--known, so that one can use observations at yet lower frequencies (where
the contrast between the CMB and the FIRB is even starker) to clean the CMB signal
out of the FIRB maps.  

To demonstrate the ability of a particular experiment to subtract the
CMB, we have forecasted errors on the determination of the FIRB power spectrum
for 4 of the \edge\ channels.  The results, in bins of $\delta l = 50$, 
are plotted in Fig.~\ref{fig:fgnds}.
For each of the error forecasts at 210, 320, 590 and 1060~GHz, we use only two
channels---the channel in question plus the 150~GHz channel (as a CMB monitor).
Note that the
590 and 1060~GHz channels require no CMB cleaning, and therefore the FIRB measurement is
sample--variance limited until the beam becomes important at $\ell > 1000$.  At 210
and 320~GHz, the errors increase above sample variance at much lower $\ell$, due
to imprecise cleaning of the CMB.  Even at these low frequencies, the CMB can
be subtracted sufficiently accurately to allow for $\delta C_l / C_l < 1$ in
bins of $\delta l = 50$ up to $\ell \sim 1000$.  The more sensitive and higher
angular resolution 100~GHz channel on \planck\ allows for even better 
CMB subtraction.  More details on these power spectrum error
forecasts are given in section 5.2.  For now, we simply state that
our calculations include the errors that result from the detector noise, sample
variance, and the cleaning of CMB using the 150 GHz channel of \edge.  They do
not include the effects of contamination by interstellar dust, which we
discuss next.

\subsection{Dust}

\begin{figurehere}
\plotone{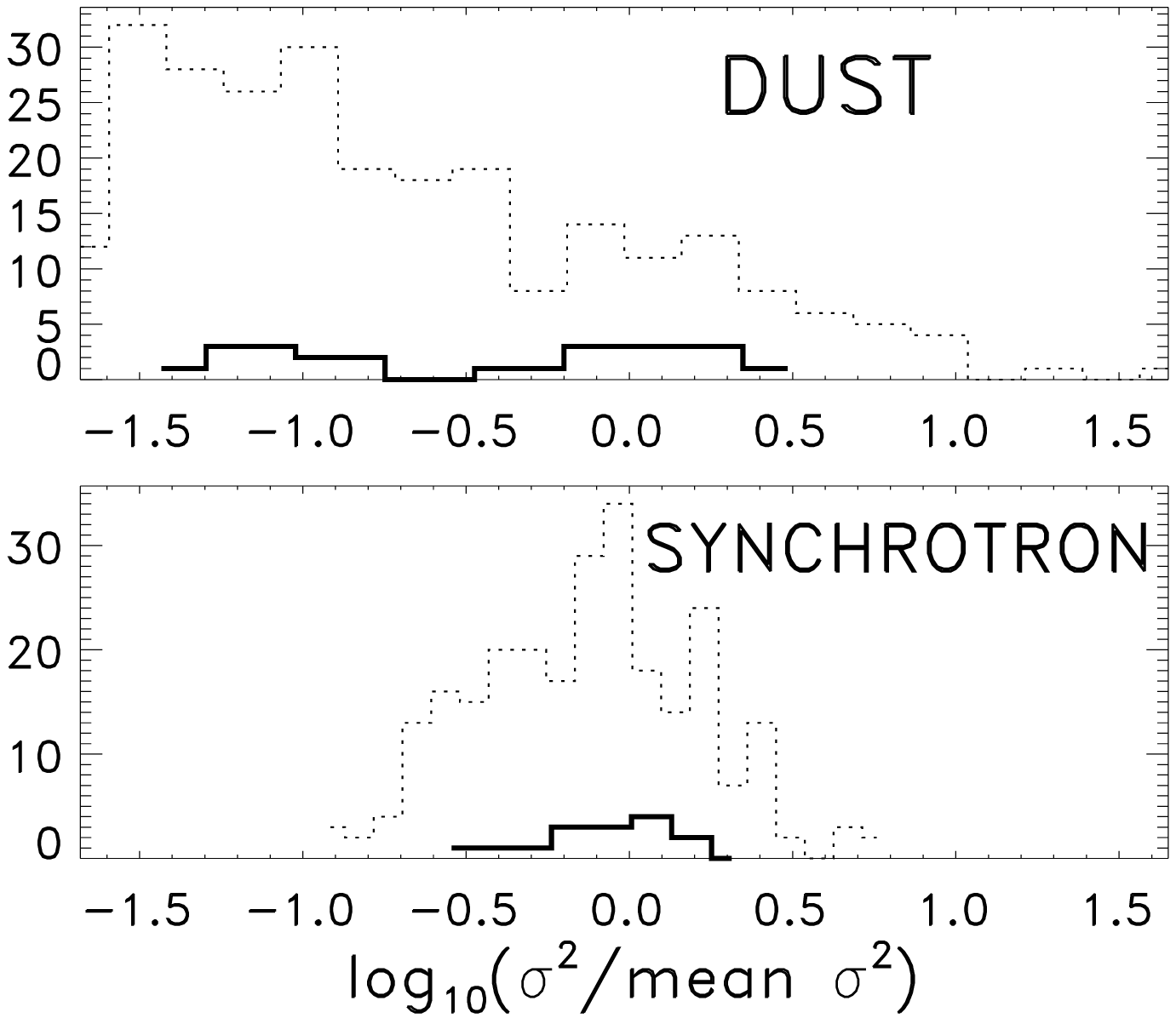}
\caption[]{\label{fig:dustrms}\protect Histograms of variance in units of the
mean variance for square patches in an $80^\circ \times 80^\circ$ field
centered on the SGP.  Light dotted lines are for the 256 patches, 
each $5^\circ \times
5^\circ$, and the heavy solid lines are for the 16 patches, 
each $20^\circ \times
20^\circ$.  The top panel is for dust emission at $300$~GHz, and the bottom
panel is for synchrotron emission at $30$~GHz.  Dust maps are the predictions
at 300~GHz \citep{finkbeiner99}.  The mean dust variances are $(9.8~\mu{\rm
K})^2$ (small patches) and $(15~ \mu{\rm K})^2$ (large patches).  Synchrotron
maps are WOMBAT predictions at 30~GHz.  The mean
synchrotron variances are $(22~ \mu{\rm K})^2$ (small patches) and $(40~
\mu{\rm K})^2$ (large patches).  } 
\end{figurehere}
%

Galactic dust is far from being an isotropic Gaussian random field and is a
more dangerous contaminant because its frequency dependence is more
uncertain and varies
spatially.  Moreover, dust can contribute significantly to the signal over the
entire frequency range where we have a hope of measuring the FIRB.  A very
encouraging point, however, is that the amplitude of the 
dust power spectrum varies considerably
over the sky.  In particular, as we will demonstrate, there are a number of
sizeable fields with substantially less fluctuation power than average, even
when that average is restricted to high galactic latitude.  Hence, an important
part of any plan to observe FIRB fluctuations is to choose fields with very low
dust fluctuation power.

We begin from the millimeter--wave predictions of the two--component dust model
of \citet[hereafter F99]{finkbeiner99}.  The two--component model has
two direction--dependent parameters which are fixed by Diffuse
Infrared Background Explorer (DIRBE) and Infrared Astronomy Satellite 
({\it IRAS}) measurements at 100 and 240~$\mu$m.  The optical properties of the
dust are assumed to be spatially uniform and were chosen to give the 
best possible agreement with the {\it COBE}/FIRAS data at longer wavelengths.
Software for making full--sky predictions of this model combined with the
{\it COBE}/DIRBE and {\it IRAS} data is available from the Wavelength--Oriented
Microwave Background Anisotropy Team\footnote{http://astron.berkeley.edu/wombat/} (WOMBAT). 

We select an
$80^\circ \times 80^\circ$ field centered on the south Galactic pole (SGP) and
calculate the power spectrum at \edge\ frequency bands in sixteen 
$20^\circ \times
20^\circ$ sub--fields.  All sub--fields had power spectra $C_\ell$ roughly
proportional to $1/\ell^3$ (as found by, e.g., \citet{gautier92,wright98}).  
The histogram of the variances of these sub--fields
is shown in Figure~\ref{fig:dustrms}.  The differences between the fields is
very large.  While the SGP as a whole is separated from the worst of the
Galaxy's dust emission, 4 of the 16 sub--fields have a variance less than 1/16
of the average SGP sub--field variance.

In Figure~\ref{fig:fgnds}, the ``Dust--AVE'' lines at each frequency are
normalized to have a variance equal to the mean variance of the SGP field as a
whole, whereas the ``Dust--LOW'' curves have a variance 1/16 smaller.  The
cleaner sub--fields have such low dust emission that the FIRB fluctuations
should completely dominate.  In light of this,
for the purposes of forecasting FIRB parameter errors in \S~5,
we have allowed ourselves to treat the dust in a very simple manner.
We assume that the dust emits a
grey--body spectrum with emissivity index $\alpha=2$ and a temperature of
$18^\circ$K, completely coherent between frequency bands, with an angular power
spectrum falling as $C_l \propto l^{-3}$.  The amplitude is the same as the
``Dust--LOW'' curves at 320~GHz.  This single--component model, normalized at
320~GHz, is sufficient for our purposes in forecasting results from \edge.  In
the low--dust regions, the dust is sufficiently unimportant that our results are
insensitive to the details of the modeling.  To fully understand what can be
learned from full--sky surveys, such as \planck, we would need a more
sophisticated modeling of the dust.  Here, we simply assume that the
``Dust--LOW'' amplitudes apply to \planck\ as well but include only 10\% of the
sky.  

In Figure~\ref{fig:fgnds}, we also show the results of repeating the
sub--division of the SGP into 256 sub--fields, each $5^\circ \times 5^\circ$.
The smaller patches show even greater variation in the dust fluctuation power.
Thus, for deep small--field observations, one can lower the dust contamination
greatly by choosing the right fields to observe.  This may be particularly
important for attempts to measure the polarization of the CMB.

Finally, for comparison, we have done the same exercise with
WOMBAT predictions for 30~GHz synchrotron maps.  These show 
significant variation, but much less than the
dust, presumably due to the larger galactic scale--height for synchrotron
emission.

\section{Forecasted Parameter Errors}
\label{sec:forecast}

With the formalism in place and the physical components specified, we can now
proceed to consider how measurements of the FIRB anisotropies constrain the
emissivity, bias, and spectral energy distributions (SEDs) of the far--infrared
sources as a function of redshift.  However, we will first describe the
degeneracy structure of the theory, so that the reader can better understand
what combinations can be constrained and how our 
quantitative investigation should proceed.

\subsection{Degeneracies and Reconstruction}

Two major degeneracies affect the interpretation of FIRB anisotropies.  First,
changes in the emissivity of the sources may be compensated by changes in their
bias (HK00), 
as evident in equation (\ref{eqn:Limber2}).  Only the product $j_b$
(eq.~[\ref{eqn:jbdef}]) can be constrained by confusion--limited anisotropy
measurements, and so we will work only with this quantity.  Of course, the
emissivity of the sources can be measured through the FIRB mean (\S 2) or
through the study of sources from higher--resolution observations.  Any
separation of bias and emissivity from anisotropy measurements will require
such external information.

Second, within confusion--limited observations, changes in the bias--weighted
emissivity are partially degenerate with changes in the spectra of the sources
coupled with a shift in redshift.  Pedagogically, this degeneracy may be best
understood by first considering the special case of a power--law matter power
spectrum, for which the degeneracy is complete.

Imagine that each redshift emits a featureless spectrum with a high--frequency
cutoff (e.g. a grey--body at some temperature).  Each redshift will contribute
anisotropy only to observed frequencies below the redshifted cutoff.  As one
considers higher frequencies, the contributions from different redshifts begin
to drop out, starting with the highest redshifts.  By measuring the reduction
in anisotropy as well as the imperfect cross--correlation, one can isolate the
contribution of the sources with that apparent cutoff frequency.  If one knows
the rest--frame frequency of the cutoff, then one can infer the redshift and use
that redshift to map the angular power of the emission to its spatial power.
By dividing the spatial power spectrum of the light by that of the mass, one
finds the bias--weighted emissivity of the sources at each redshift.

However, if the emitted temperature is unknown, one cannot assign a redshift to
a particular observed cutoff.  For example, an increase in temperature of the
emission would imply a larger redshift.  This shift in redshift alters the
mapping between physical and angular scale; however, because the spatial power
spectrum is a power--law, the shift in distance scale can be compensated by a
shift in normalization so as to leave the observed angular power spectrum
unchanged.  We thus have a degeneracy between the emitted spectrum, the
redshift, and the bias--weighted emissivity of the sources.  In models of
interest, a 10\% shift in $1+z$ mimics a 30\% shift in fluctuation amplitude
(60\% in power).

Another way to describe this is to view the FIRB anisotropies as a sum of
correlations of different {\it apparent} temperature components.  By analyzing
the frequency structure of the correlations, one can determine the power
spectrum associated with each apparent temperature.  However, unless one knows
the original emitted temperature, one cannot determine the redshifts of the
apparent temperature components.  Without the redshift, one cannot convert the
angular power spectrum to a spatial one.

The degeneracy is lifted when the power spectrum is not a power--law.  With a
feature or bend in the spatial power spectrum, one has a second handle on the
redshift of the sources through the measurement of where the feature appears in
angle.  By combining the details of the frequency covariance with the location
of features in the angular power spectrum, one can measure the emission
spectrum separately from the bias and luminosity.  Unfortunately, because CDM
power spectra are fairly smooth and lack strong features, the
redshift--temperature degeneracy is only partially lifted.  The slow dependence
of the comoving angular diameter distance on redshift further limits one's
ability to measure the redshift.

The above paragraph assumed that the bias of the sources was scale--independent,
so that scales in the matter power spectrum would shine through in the FIRB
anisotropies.  The bias may instead be scale--dependent.  Even a smooth
dependence on scale can shift the location of soft bends in the power spectrum,
such as are found in CDM models.  This would confuse the interpretation of the
angular power spectrum in terms of redshift and skew the inferred bias--weighted
emissivities.  Of course, a general scale--dependent bias will ruin any use of
the angular power spectrum to measure the redshift of the sources.

Because of the above concerns, we anticipate that the emission temperature and
hence redshift of the sources will not be well--constrained by the FIRB
anisotropies alone.  In \S \ref{sec:quant}, we find that biases could be
constrained in 4 redshift bins at the level of 5--15\% 
if the emission spectra were known, but the
large--angle FIRB anisotropies themselves will not deliver the $\sim\!5$\%
emission temperature constraint that would be needed to break the degeneracy to
this level.  Instead, one must rely on external information, notably from
high--resolution infrared imaging with followup spectroscopy, to supply the
emission spectrum as a function of redshift.  Clearly, the spectral energy
distribution of bright sources will be a good start.  We caution only that the
FIRB anisotropies actually require the bias-- and luminosity--weighted spectral
energy distribution of the sources.  If the spectra of sources depends strongly
on luminosity, it may be necessary to extrapolate to unresolved flux levels
when integrating over the luminosity function.

Having noted this degeneracy, it is important not to over--react to it.  First,
allowing for uncertainty in the emission spectra will make the bias--weighted
emission $j_b$ uncertain, but there will be a combination of these parameters
that will be well--constrained.  Indeed, as shown in Appendix A of
\citet{eisenstein99}, the errors on the product of $j_b$ and a calculable
function of the spectral variation will be the same as the errors on $j_b$ in
the limit that the spectra are held fixed.  In our quantitative work, we want
to focus on this well--constrained combination of parameters; in breaking the
degeneracy by fiat, we can calculate the errors on this quantity.  We will show
in \S \ref{sec:quant} that the FIRB anisotropies can deliver several 
better than 10\% constraints on the properties of FIRB sources.

Second, the emission temperatures in plausible models do not vary by
large amounts.  If the FIRB anisotropies were to reveal a factor of
$8$ change in emission as a function of redshift, it would be
unrealistic to explain it by a factor of two shift in the temperature
scale.  While such coarse statements ``waste'' the precision of the
FIRB anisotropy measurement, they do show that the measurements can
yield interesting results without detailed external constraints on the
emission properties.

In the same spirit, we will assume that the underlying cosmology is well
known.  Introducing uncertainties in that sector would cause additional
degeneracies.  Some are simple: the amplitude, shape, or time dependence
of the matter power spectrum will be directly degenerate with the bias.
Uncertainties in the relations between distance, volume, and redshift
will cause more subtle problems akin to those described above.  However,
one can always map the constraints on $j_b$ into constraints on a product
of the $j_b$ and functions of cosmology.

\subsection{A Quantitative Study}
\label{sec:quant}

For a quantitative assessment of what we can learn about $j_b(k,\nu,z)$ from
measurement of $C_l^{\nu \nu'}$, we split $j_b$ into four redshift bins with
intervals of [0,0.5], [0.5,1], [1,2] and [2,4].  We then parameterize $j_b$ as
follows:
\be
\label{eqn:parametrize}
j_b(k,\nu,z)=\sum_i \chi_{i(z)} \alpha_i j^s_b(k,\nu,z)(k/k_*)^{n_i}
\ee
where $\chi_{i(z)}$ is unity for $z$ in the $i^{\rm th}$ redshift bin and zero
otherwise.  The parameters $\alpha_i$ and $n_i$ adjust the amplitude and
scale--dependence, respectively, of the bias--weighted emissivity in the $i^{\rm
th}$ redshift bin.  $j^s_b(k,\nu,z)$ is simply the $j_b$ of the HK00 model
(and actually has no $k$ dependence).
The pivot point $k_*$ is chosen to project to $\ell=500$ at all redshifts.  As
we will see, this is nearly the ``sweet spot'' at which the errors on 
$\alpha_i$ and $n_i$ become independent.  In total, the model 
has 8 free parameters: 4 amplitudes and 4 tilts.

\begin{table*}[hbt]\small
\caption{\label{table:channels}}
\begin{center}
{\sc\edge\ Channel Characteristics}\\
\begin{tabular}{ccccccccc}
\tableskip\hline\hline\tableskip
Channel & $\nu$ & $\lambda$ & Beam & Number of & NET$_{\rm RJ}$ &
\multicolumn{2}{c}{$\sigma_{\rm pix}$} & $N(>5\sigma_{\rm pix})$ \\
 & & & FWHM & Detectors & per detector & & &\\
& (GHz) & ($\mu$m) & ($^\prime$)& &($\mu$K\, $\sqrt{\rm s}$) & ($\mu$K) &
(mJy) &\\ 
\tableskip\hline\tableskip
L1 & 150 & 2000 & 14 & 1   &   68 & 6.2& 79& $< 1$\\
L2 & 220 & 1360 & 14 & 1   &   36 & 3.3& 76&$< 1$\\
H1 & 320 & 940 & 6 &6&  91 & 9.8 &99&7\\
H2 & 390 & 770 & 6 &6&  71 & 7.7  &114&12\\
H3 & 480 & 630 & 6 &6&  60 & 6.5&143&21\\
H4 & 590 & 510 & 6 &6&  49 & 5.3&176&30\\
H5 & 710 & 420 & 6 &6&  42 & 4.6 &227&33\\
H6 & 870 & 340 & 6 &6&  36 & 3.9 &274&38\\
H7 & 1,060 & 280 & 6&6& 30 &  3.3&352&40\\
H8 & 1,290 & 230 & 6&6& 26 &  2.8&454&39\\
\tableskip\hline
\end{tabular}\\[12pt]
\begin{minipage}{5.2in}
NOTES.---%
Pixel errors $\sigma_{\rm pix}$ are for beam--size pixels and observations
uniformly covering 1\% of the sky over 10 days.  The weight--per--unit
solid angle of equation (\ref{eqn:littlec}) is given by
$w = 1/(\sigma_{\rm pix}^2 {\rm FWHM}^2)$.  For the H
(``high--frequency'') channels, we assume only 4 of the six detectors
are used.  Number of sources observed $N(>5\sigma_{\rm pix})$ assumes
model E (G98).  See http://topweb.gsfc.nasa.gov for more details.
\end{minipage}
\end{center}
\end{table*}

We can forecast errors on these parameters by theoretically propagating
the measurement errors.  This is conveniently done with the Fisher 
matrix formalism \citep{kendall69,tegmark97b}.  
In general, the Fisher matrix for a set of parameters, $a_p$, depends on the
covariance matrix of the data (in this case the maps) and the derivatives of
the covariance matrix with respect to the parameters:
\be
\label{eqn:fisher1}
F_{pp'} = {1\over 2} {\rm Tr}\left[C^{-1} {\partial C \over \partial a_p}
C^{-1} {\partial C \over \partial a_{p'}}\right]
\ee
where the trace here runs over the suppressed indices $\ell$, $m$ and $\nu$.  
Calculating the Fisher matrix is most easily done in spherical harmonic space
where the covariance matrix of the maps can be written as
\be
C_{\nu l m,\nu' l' m'} \equiv \langle a_{\nu l m} a_{\nu' l' m'} \rangle
= c_l^{\nu \nu'} \delta_{ll'}\delta_{mm'}
\ee
with
\be
\label{eqn:littlec}
c_l^{\nu \nu'} \equiv \sum_i C_l^{\nu \nu' (i)}
b_\nu(l) b_{\nu'}(l) + w^{-1}(\nu) \delta_{\nu \nu'}
\ee
where $b_\nu(l)$ is the window
function of the beam, $w(\nu)$ is the weight--per--unit solid angle of
the map at frequency $\nu$, and the index $i$ labels the various
components, which we have assumed to be uncorrelated.
Such multi--frequency, multi--component analyses have been performed for CMB 
experiments by a number of authors 
\citep{tegmark96,bouchet96,knox99,tegmark00,cooray00}.  
We use four components in our modeling: the FIRB correlations, the
FIRB shot noise, Milky Way dust, and the CMB.  We take the shot noise
to be uncorrelated between different frequencies for the sake of
simplicity.  In reality, the cross--frequency correlations are probably
strong.  We justify our approximation below.

We can separate the calculation into separate multipoles, so that
with uniform coverage over a fraction of the sky, $f_{\rm sky}$,
\be
\label{eqn:fisher2}
F_{pp'} = \sum_l {\left(2l+1\right)f_{\rm sky} \over 2}
{\rm Tr}\left[c^{-1}_l {\partial c_l \over \partial a_p}
c^{-1}_l {\partial c_l \over \partial a_{p'}}\right]
\ee
where now all the terms in the trace are matrices with (suppressed) indices
$\nu,\nu'$.  When $f_{\rm sky}$ is less than 1, this equation is
approximate, although it is generally a very good approximation.  
The $f_{\rm sky}$ factor
accounts for the decreased number of modes available when the sky
coverage is less than full.   

At this point, the matrices one needs to invert are only size
$n_{\rm ch}$ by $n_{\rm ch}$, where $n_{\rm ch}$ is the number of channels, and
therefore this calculation can be done quickly.  To simplify 
the amplitude derivatives, we take them with respect to 
$\alpha_i^2$ instead of $\alpha_i$.

The diagonal element $F_{ii}$ gives the variance on the $i^{\rm th}$ parameter
when all the others are being held fixed (the {\it unmarginalized} case), while
the inverse of the Fisher matrix yields the predicted covariance matrix when
all parameters are allowed to vary simultaneously (the {\it marginalized}
case).  We will also display the results when the amplitudes are allowed to
vary but the scale--dependence parameters are held fixed.  
For all our calculations we 
restrict the sum in equation (\ref{eqn:fisher2}) to $\ell < 1600$ in order to reduce
the dependence of our results on the amplitude of the shot--noise power
spectra, which are quite uncertain.

We begin by displaying the predictions for the \edge\ mission in
Figure~\ref{fig:error_edge}.  In the unmarginalized case, the two parameters
of the 3 low--redshift bins can be constrained to $\sim\!1\%$, while the
highest--redshift bin is measured only to $\sim\!5\%$.  In the marginalized
case, all parameters are constrained at the $\sim\!10\%$ level.  Interestingly,
the constraints on the amplitude are very similar whether or not the tilts
$n_i$ are allowed to vary.  This indicates that the values $k_*$ were
well--chosen so that the uncertainties in the amplitudes and tilts 
are nearly uncorrelated \citep[for more details, see]{eisenstein99}.

Since \edge\ is covering $1\%$ of the sky to $\ell\approx1000$, one would have
expected it to yield a $1/\sqrt{0.01\times 1000^2}=1\%$ measurement of some
combination of these parameters.  Indeed, diagonalization of the $4\times4$
amplitude sub--block of the Fisher matrix shows that this is the case.
Table~\ref{table:eigenpairs} shows that the errors in the fully marginalized case
displayed in Figure~\ref{fig:error_edge} are dominated by one eigenvalue.
Excursions in parameter space that move all of the $j_b$ amplitudes in the same
direction are constrained at better than 1\%, whereas excursions that move
neighboring redshift bins in opposite directions are only constrained to 10\%.
Indeed, our choice to use only 4 bins constitutes an implicit smoothing prior
on the set of all possible excursions of the function $j_b(\nu,z)$, and because
the FIRB correlations are projected quantities, one will never constrain highly
oscillatory star--formation histories.  Physically, of course, such models are
absurd.  Models that differ by broad, smooth changes can be distinguished by
\edge\ at high accuracy, $\sim\!1\%$, which is better than the marginalized
errors in Figure~\ref{fig:error_edge} would indicate.


\begin{figurehere}
\plotone{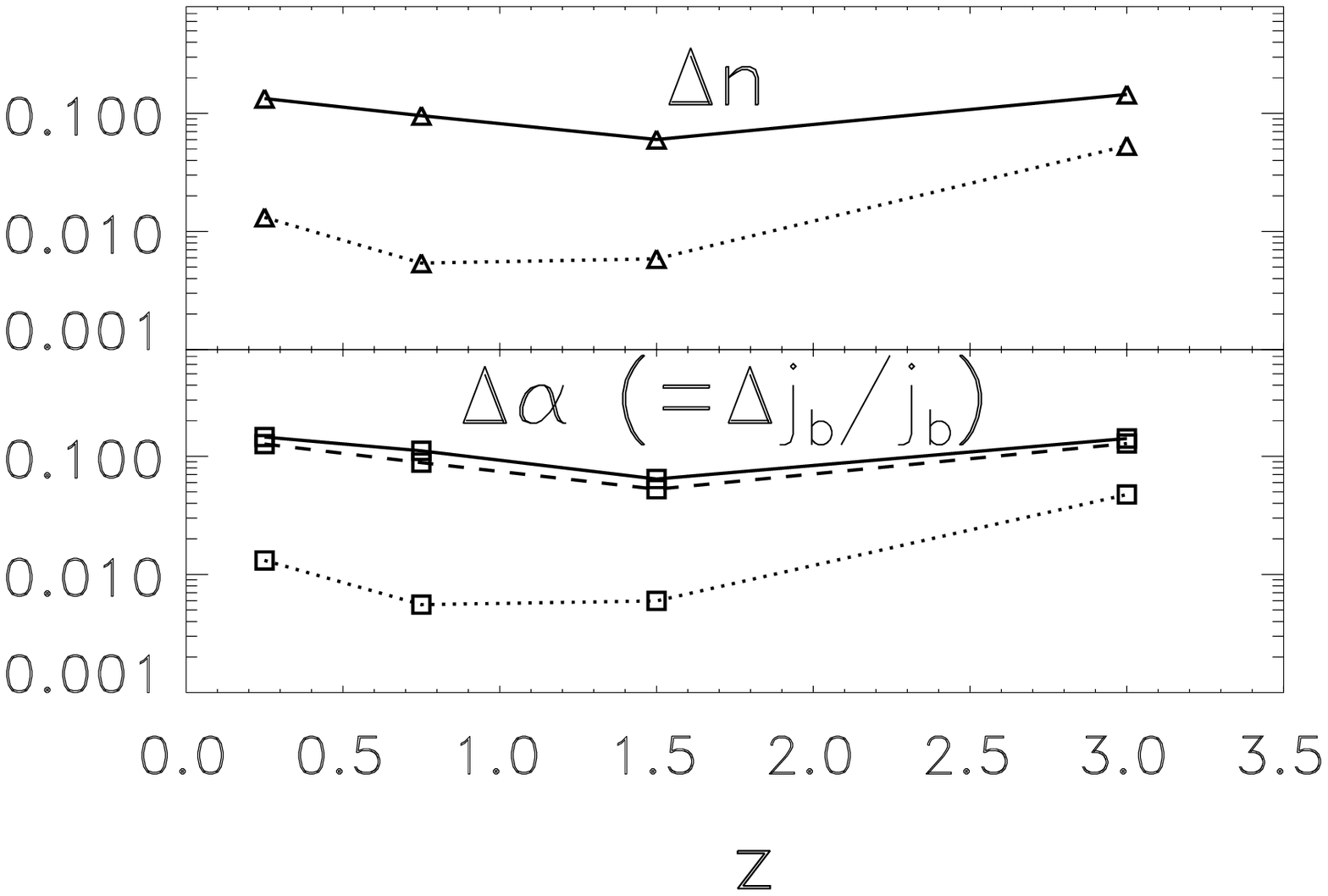}
\caption[]{\label{fig:error_edge} Forecasted relative errors from \edge\
observations on the $j_b$ amplitudes (bottom panel, squares) and absolute
errors on the bias spectral index parameters (top panel, triangles) for a model
with 8 parameters: the amplitude factor $\alpha$, and $n$ in 4 redshift bins. 
Dotted lines connect
results for parameter errors assuming the other 7 parameter held fixed.  Solid
lines connect the results for parameter errors that follow from assuming none
of the parameters are fixed.  The dashed line shows the result for the 
amplitude parameters when holding only the $n$ parameters fixed.  }
\end{figurehere}
%

We repeat all of the analysis for the \planck\ mission.  Since \planck\ covers
the full sky, over much of which the dust is a significant contaminant, an
accurate treatment of how well \planck\ can measure the $\alpha$ and $n$
parameters of equation (\ref{eqn:parametrize}) would require a more careful 
treatment of the dust contamination than
has been done here.  To avoid these complications, we forecast the results for
a conservative analysis of the \planck\ data which only uses the cleanest 10\%
of the sky.  We assume that the fluctuation power in this cleanest 10\% is the
same as assumed for the \edge\ observations of 1\% of the sky.  

The results for \planck\ are shown in Figure~\ref{fig:error_hfi} and
Table~\ref{table:eigenpairs}.  The behavior is quite similar.  The difference
between the errors on $\alpha_i$ with and without marginalizing over the $n_i$
are larger in the \planck\ case.  This indicates that the $\alpha_i$ and $n_i$
are somewhat correlated.  A slightly larger choice for $k_*$ would have
decorrelated the two sets of parameters, and the errors on $\alpha_i$ would
have been simply those in the case where the $n_i$ were not varied.  In other
words, with a better choice of $k_*$ the amplitude constraints would be 
quoted as closer to 5\% than 10\%.
The movement of the sweet spot to higher $k$ is a direct result of the 
greater sensitivity and slightly higher angular resolution of \planck.

\begin{figurehere}
\plotone{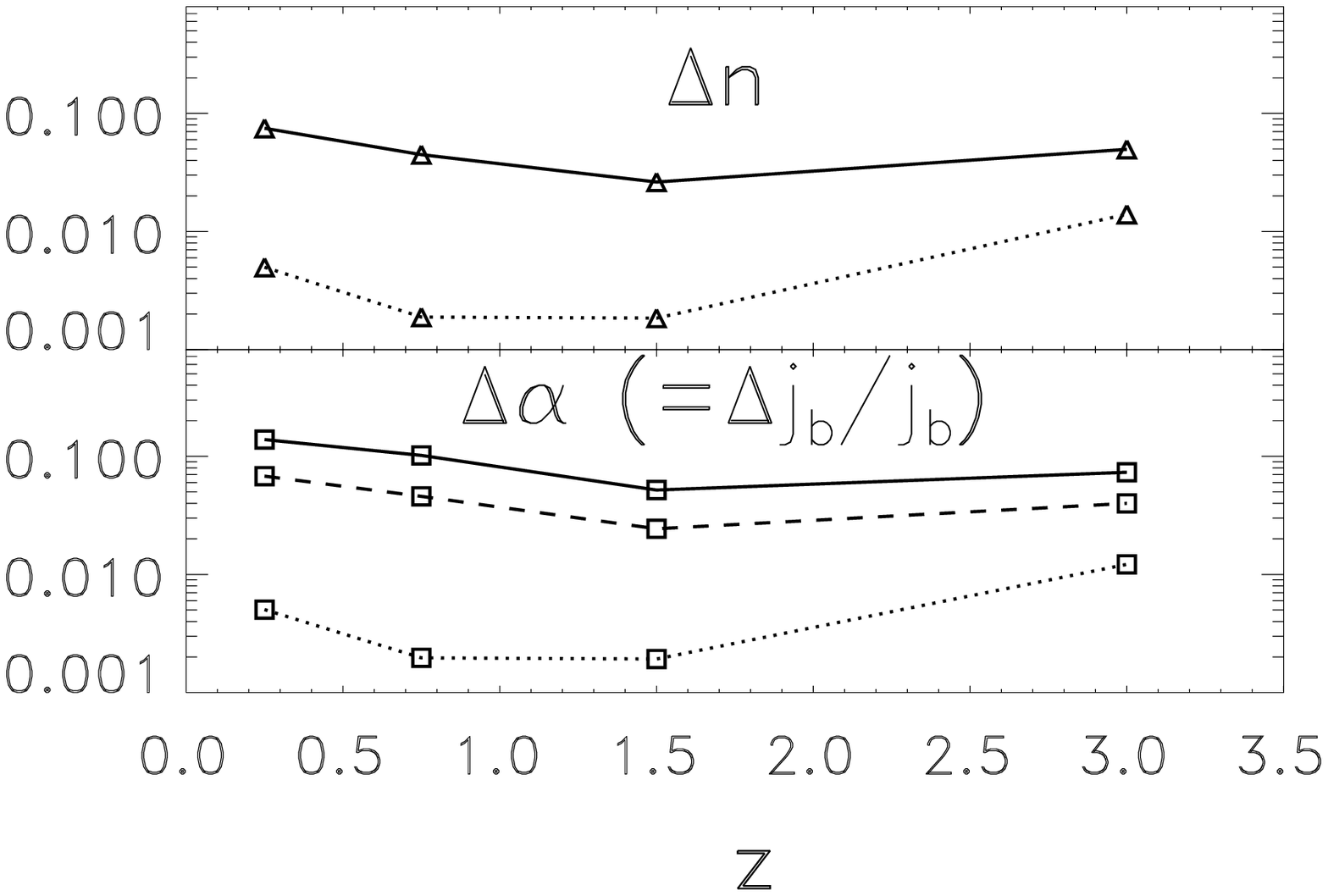}
\caption[]{\label{fig:error_hfi}
Same as Figure~\ref{fig:error_edge} but for \planck. 
} 
\end{figurehere}
%

\begin{table*}[bt]\small
\caption{\label{table:eigenpairs}}
\begin{center}
{\sc Diagonalizing the Fisher Matrix}\\
\begin{tabular}{cccccc}
\tableskip\hline\hline\tableskip
Experiment & Error on bias & \multicolumn{4}{c}{Eigenvector} \\
& $\lambda_F^{-1/2}/2$ & $0<z<0.5$ & $0.5<z<1$ & $1<z<2$ & $2<z<4$ \\
\tableskip\hline\tableskip
\edge   & 0.0040 & 0.2799 &  0.7094 &  0.6451 &  0.0471 \\
        & 0.0151 & 0.4525 &  0.4860 & --0.7148 & --0.2192 \\
        & 0.0777 & 0.5798 & --0.2458 & --0.0380 &  0.7759 \\
        & 0.1922 & 0.6171 & --0.4474 &  0.2673 & --0.5896 \\
\tableskip\hline\tableskip
\planck & 0.0014 & 0.2508 &  0.6780 &  0.6884 &  0.0588 \\
        & 0.0060 & 0.4097 &  0.5361 & --0.6469 & --0.3554 \\
        & 0.0234 & 0.4801 & --0.0460 & --0.2023 &  0.8523 \\
        & 0.0912 & 0.7340 & --0.5008 &  0.2582 & --0.3793 \\
\tableskip\hline
\end{tabular}\\[12pt]
\begin{minipage}{4.6in}
NOTES.---%
The eigenvalues and eigenvectors of the Fisher matrices for \edge\ and
\planck.  The scale--dependence parameters $n_i$ have been held fixed
here, leaving only the four amplitude parameters $\alpha_i$.
Each row gives an eigenvalue and the corresponding eigenvector.
The eigenvalues are listed as the inverse square
root of the Fisher matrix eigenvalue $\lambda_F$ divided by two; this is
the 1-$\sigma$ error on the bias.  The factor of two comes from our
use of $\alpha_i^2$ as a parameter instead of $\alpha_i$.  
The eigenvectors specify how that error is
divided between the four redshift bins.  Note that the spectrum
of eigenvalues increases rapidly: model changes that would move all
of the biases up or down together are extremely well constrained
(better than 1\%),
whereas changes that shift the biases in alternating fashion are 
relatively poorly constrained (10-20\%).
\end{minipage}
\end{center}
\end{table*}

We have studied the dependence of the
parameter errors to changes in experimental parameters and 
the parameters governing sources of astrophysical noise (namely, 
the shot noise and dust power spectrum amplitudes).
Halving the dust power spectrum makes less than 10\% changes in the errors.
Increasing it by a factor of 20 makes at most a factor of 2
increase in the parameter errors.  Increasing
the shot noise power spectrum by a factor of 2 makes less
than a 10\% change in the errors.

This robustness is fortunate since, as has already been
mentioned, our modeling of these contaminants
has not been very sophisticated.  For the dust we have
assumed complete coherence, and for the shot noise complete
decoherence; reality for both is somewhere in between.
In principle the $a_p$ of equation (\ref{eqn:fisher1}) would be not 
only the FIRB parameters,
but also dust and shot--noise parameters.
In effect, our calculation assumes that the $C_l^{\nu \nu'}$ for the
dust and shot noise are perfectly known.  
In reality these statistical properties will have to be
determined from the data, although external datasets will
provide useful information as well.  
A further weakness of
our dust modeling is the implicit assumption of statistical
isotropy.  A more rigorous treatment of the dust, which one
would need to study FIRB anisotropy in regions of typical
dust fluctuation power, would be quite challenging indeed.

Removal of low frequency channels degrades
our high--$z$ constraints, as we would expect.  
Eliminating the \planck\ 217~GHz channel,
while not changing the unmarginalized errors, does increase
the marginalized errors---in the lowest--$z$ bin by 50\% and the
highest--$z$ bin by a factor of 2.  Removing the three lowest
frequency FIRB channels (L2, H1 and H2) has a similar effect.  
However, simply removing channel L2 makes almost no difference.
Presumably the importance of LS would be more readily apparent
if the highest--$z$ bin,
which runs from $z = 2$ to $z=4$, were split up more finely.
As the bin stands, the H1 and H2 channels are helping to constrain its
amplitude because of their sensitivity to just the lower end of the 
range.

The high frequency channels are important
for separating out the low--redshift contribution from the
high--redshift contribution.  
Removing the highest three frequency channels of \edge\
has only a small effect on the unmarginalized errors, but the
marginalized errors increase by factors of $\sim 10$.
For \planck\, removal of the 857~GHz channel results in mostly
small changes to the unmarginalized errors 
but factors of from 4 to 8 increase in the marginalized errors.

While frequency range is important, we have not been able to see, with
this study, a benefit from a dense sampling of that range, such as
offered by \edge.  Removing all the even \edge\ channels (L2, H2, ...,
H8) while simultaneously doubling the weight of the odd ones (to keep
the total weight roughly unchanged) results in practically no change
in either the marginalized or unmarginalized errors.  This may
be due to the fact that our bins of $1+z$ are coarser than
the frequency bins; i.e., the finer the frequency sampling, 
the faster the changes in emissivity with redshift 
one should be able to detect.  Dense sampling also makes possible
more consistency tests and better monitoring of possible dust
contamination.

Once again we stress that our results here have assumed small
departures around a particular model for the FIRB fluctuations.
In particular, it may be possible to have detections of significant
emissivity at redshifts beyond 4, if there is such emission.  Even
if there is not such extra emission, and $\Delta j_b / j_b$ turns out
to be large for $z > 4$, one may still be able to set very interesting
upper limits on $j_b$.  Combined with a lower limit on the bias,
this could be turned into an upper limit on $j_\nu$ alone---providing
a new constraint on energy production at high--redshift.

It is now convenient to give some more details on the FIRB power spectrum
error forecasts in figure 6.  They were the same as the parameter
error forecasts described in this section (i.e, equation (16) through 
equation (19)) 
with the following changes:
(1) for the errors in each panel only two channels were used---the one
corresponding to that panel, and the 150~GHz channel as a CMB monitor,
(2) the parameters were the FIRB power spectrum multipole moments
themselves, $C_l^{\nu \nu}$, instead of the $\alpha_i$ and $n_i$,
(3) dust was ignored, and (4) the FIRB $C_l^{\nu \nu}$ 
included contributions from non-linear corrections.
It should be clear that the figure 6 error forecasts were only used for that
figure.  They are not an intermediate step in our calculation of FIRB
parameter errors.

\section{Discussion}
\label{sec:discussion}

The large--angle anisotropies of the FIRB offer a complementary view of the
high--redshift universe to that given by direct investigation of far--infrared
sources.  Clearly, if one could measure the redshift and spectrum of every
far--infrared source in some large patch of sky, then one would recover all of
the information we have described and more.  However, this will not be possible
in the near future.  By combining high--resolution measurements of the average
spectrum and emissivity as a function of redshift with the known statistics of
large--scale density fluctuations, the confusion--limited measurement of the FIRB
anisotropies gives us a window on the large--scale biasing of the far--infrared
sources.  In CDM cosmologies, this bias holds implications for the typical mass
of the host halo of the sources \citep{mowhite96}, thereby giving an
observational window on the halo environment of the starburst and AGN activity
that power the FIRB.  Theories for how the galaxy density--morphology relation
arises \citep{kauffmann93} and why clusters of galaxies form rather than single
overmerged galaxies often appeal to there being a preferred mass (or velocity)
scale for galaxy interactions \citep{kolatt99,somerville00}.  If such
interactions lead to dust--obscured star formation and nuclear activity, then
the bias of the FIRB sources could inform the modeling of bulge and group
formation.

It is interesting to compare the limits on clustering that can be gained from
confusion--limited measurements to those that could be gained from more typical
correlation analyses of sets of individual sources.

First, we will consider the case in which exact redshifts for the sources are
known and ask what number and density of sources would be needed to measure the
power spectrum over a broad band centered at a comoving 
wavenumber $k=0.2h\Mpc^{-1}$ ($\ell\approx800$) to
We assume that $P(k) = 1000h^{-3} \Mpc^3$ at $k=0.2h \Mpc^{-1}$; this
corresponds to a bias of 2 at $z\approx2.5$.  For a fixed number of objects
$N$, the power spectrum error bars are minimized when the comoving volume
density is equal to $P^{-1}$ \citep{kaiser86,tegmark97a}; in this case, the
fractional error is approximately 
$2\sqrt{2}/\sqrt{N {k^3 P(k) \over 2\pi^2}}$.
Hence, 5\% accuracy
requires approximately 8000 sources over a comoving volume of $8\times 10^6
h^{-3} \Mpc^3$.
\footnote{Of course, the required number and volume scale as the inverse 
square of
the desired accuracy; 1\% accuracy would require $\sim 200,000$ sources.}.  
If this volume is roughly cubic, then this is
approximately a range of 0.25 in redshift and 8 square degrees at $z=2.5$.  
If 
the source density is much smaller or larger than $10^{-3}h^3 \Mpc^{-3}$, then
the number of sources required to achieve a 5\% measurement of large--scale
power increases dramatically.

Next, we consider the case in which the redshifts are only approximately known,
e.g.~by using the far--infrared or radio spectral index
\citep{carilli99,blain99e}.  
We approximate this situation by saying that the sources can
be divided into a series of redshift slices.  Each slice is analyzed by angular
correlation methods and is presumed to be statistically independent from the
others.  In the limit that the conformal distance $r$ changes only slightly
across the redshift slice, the angular power spectrum $C_\ell$ of the sources
in the slice is simply the spatial power spectrum at $k=\ell/r$ divided by the
comoving volume per steradian in the slice.  
For $P=1000h^{-3}\Mpc^3$ and a
slice of unit redshift at $z=2.5$, this gives $C_\ell\approx10^{-7}$.  
The optimal number
of sources per steradian to minimize the errors on the measurement of $C_\ell$
for a given number of sources is simply $C_\ell^{-1}$.  With this surface
density, the fractional errors on $C_\ell$ are roughly 
$2\sqrt{2}/\sqrt{N {\ell^2 C_\ell \over 2\pi}}$.
Inserting $\ell=800$, one finds a requirement of roughly
$3\times10^5$ sources (in the unit redshift band) spread over 100 square
degrees in order to achieve 5\% fractional errors on the power spectrum.

In both cases, the amplitude of the power spectrum gives a preferred surface
density in a narrow redshift band.  This density implies an upper limit to the
flux cut of the selection.  If we assume that the comoving source density is
constant out to $z=5$ at this flux level, then integrating over redshift
implies $\sim$15,000 total sources per square degree.  
This corresponds to
roughly 1~mJy at 850~$\mu$m \citep{hughes98}.  One gets the same answer in
either of the two cases above and regardless of the thickness of the redshift
slice in the angular analysis; the value is controlled by the amount of volume
at high--redshift and the amplitude of the power spectrum one is trying 
to measure.
The detection of sources becomes confusion--limited when the inverse of the
source surface density is less than 30 times the beam area.  Hence, one can
immediately calculate that detecting sources at the density needed to maximize
one's measurement of the high--redshift power spectrum with a fixed number of
sources requires a beam smaller than 5".  Surveys with beams larger than this
will reach their confusion limit before achieving the optimal source density
and will need to measure more sources in order to gain the same fractional
limits on the power spectrum.  Surveys with beams smaller than this should not
integrate to the confusion--limit but instead cover more sky.

Clearly, resolving the FIRB sources has other benefits with respect to
clustering that cannot be matched by a confusion--limited measurement; for
example, one can study the clustering of different populations of sources.
However, the above calculations show that achieving accuracy equivalent to
Figure \ref{fig:error_edge} on the large--scale power spectrum requires large
sets of sources: hundreds of thousands with crude redshifts or many thousand
with accurate redshifts.

An aspect of the FIRB anisotropies that we have not included in our treatment
is the possibility of stochastic bias \citep{dekel99}.  Our assumption that the
bias between the emission at different frequencies is perfectly correlated
(i.e. that the frequency covariances can be decomposed into $b(\nu)b(\nu')$) is
optimistic because it allows measurements of the cross--correlations between
frequencies to constrain the $j_b$ directly.  However, we find that the HK00
predictions of the cross--correlation differ sufficiently little from perfect
correlations that the leverage on the redshift decomposition is relatively
weak.  Breaking the assumption would slightly weaken our constraints but would
also open up a new sector of physical parameters, namely the stochasticity of
the relative bias between different frequencies.  This can occur if at a given
redshift, there are multiple components with differing spectral energy
distributions that are imperfectly correlated with each other (and thus
necessarily with the density).  Such flawed correlations are necessary in the
non-linear regime, where the density contrast can't fall below $-1$, but it is
unknown whether bias is stochastic on the large scales studied here.  If the
observed correlation coefficients were 
more distinct from unity than the baseline
predictions, the lack of correlation could be measured and perhaps interpreted.
Obviously, it would be intriguing to associate the separate populations with
active nuclei versus star formation or quiescent star formation versus
starbursts.  Hence, the cross-correlations between frequencies could be an
interesting route for FIRB anisotropies to constrain the properties of
high-redshift sources.

There remains the possibility of truly diffuse emission contributing to
the FIRB, i.e., emission which does not resolve into distinct sources.
As an example, grey dust distributed throughout the IGM has been
invoked as a non-$\Lambda$ explanation for the anomalously dim 
high--$z$ supernovae \citep{aguirre99}, and it has been shown that if
this explanation is correct, then a significant contribution to the
FIRB at low frequencies comes from this grey dust \citep{aguirre00}.
We can expect the clustering properties of such emission to be
different from that of the individual sources.  A cross-correlation
analysis of a wide-field map of resolved FIRB sources with a map of
the unresolved FIRB emission (as measured by, e.g., \planck\ or \edge)
could eventually be used as a probe of any such diffuse component.

\section{Conclusions}

We have presented a study of the large--angle clustering properties of
the far--infrared background, focusing on the role of confusion--limited
anisotropy experiments.  We find that under reasonable assumptions,
and in regions of the sky with unusually low Galactic dust emission,
correlated FIRB fluctuations should dominate the observed 
angular power spectrum over a wide
range of frequencies and angular scales.  As a result, such experiments
will provide useful constraints on any model for the sources of the FIRB.
The main drawback of the confusion--limited technique is that it cannot
easily separate different physical classes of sources, and so many
important questions about the nature and distribution of far-infrared
sources will remain squarely in the domain of high--resolution studies.
However, matching the constraints on the large--scale clustering that
could be produced by the \edge\ or \planck\ experiments requires an
impressively large set of point source measurements with or without
accurate redshift information.

To model the FIRB fluctuations, one must specify the spectrum, emissivity,
and bias of high--redshift galaxies and AGN, as well as the redshift
evolution of these quantities.  Despite the inherent degeneracies, we find
that confusion--limited anisotropy observations yield tight constraints
on certain combinations of these model ingredients.  Without external
spectral information, the emissivity is degenerate with the temperature of
the sources.  However, as the plausible ranges of temperatures and biases
are reasonably small, the FIRB anisotropies by themselves could detect
strong variations in the emissivity as a function of redshift, notably the
turn--on of embedded star formation at high redshift.
We find that if the
mean emission spectrum was precisely known, the bias--weighted emissivity
could be measured to $\sim\!10\%$; while strong priors on the amount of
oscillation allowed as a function of redshift can allow these constraints
to approach 1\%.  These constraints are tighter than differences between
possible models for the sources of the FIRB, rendering such studies of
the infrared sky a useful probe of the high--redshift universe.

\acknowledgments

We thank S. Meyer, W. Hu, and M. Zaldarriaga for many
useful conversations, J. L. Puget for pointing out how much the dust power
spectrum varies spatially, and D. Finkbeiner for assistance with the 
WOMBAT dust and synchrotron predictions. 
LK is supported by the DOE, NASA grant NAG5-7986
and NSF grant OPP-8920223.  DJE and ZH were supported by NASA through Hubble
Fellowship grants \#HF-01118.01-99A and \#HF-01119.01-99A from the Space
Telescope Science Institute, which is operated by the Association of
Universities for Research in Astronomy, Inc, under NASA contract NAS5-26555.
AC acknowledges support from J. Carlstrom, W. Hu, and D. York.

\end{document}